\documentclass[traditabstract]{aa}
\usepackage[pdftex]{graphicx}
\usepackage{txfonts}
\usepackage{url}
\usepackage{natbib}
\usepackage{aas_macros}
\usepackage{placeins}

\newcommand{\lya}{Ly-$\alpha$}

\begin{document}

\title{Statistics and characteristics of MgII absorbers along GRB lines of sight observed with VLT-UVES}

 \author{Susanna D. Vergani\inst{1,2,3}
        \and
        Patrick Petitjean\inst{2}
        \and
        C\'edric Ledoux\inst{4}
        \and
        Paul Vreeswijk\inst{5}
        \and
        Alain Smette\inst{4}
        \and
        Evert J.A. Meurs\inst{3,6}}  

\institute{University Paris 7, APC, Lab. Astroparticule et Cosmologie, UMR7164 CNRS, 10 rue Alice Domon et Lonie Duquet, F-75205, Paris Cedex 13, France 
\and
          University Paris 6, Institut d'Astrophysique de Paris,
          UMR7095 CNRS, 98bis Boulevard Arago, F-75014, Paris, France
            \and
  School of Physical Sciences and NCPST, Dublin City University, Dublin 9, Ireland
          \and 
          European Southern Observatory, Alonso de C\'ordova
          3107, Casilla 19001, Vitacura, Santiago 19, Chile
          \and        
          Dark Cosmology Centre, Niels Bohr Institute, University of Copenhagen, DK-2100 Copenhagen, Denmark
\and
School of Cosmic Physiscs, DIAS, 31 Fitzwilliam Street, Dublin 2, Ireland
       }
\offprints{S.D. Vergani, vergani@apc.univ-paris7.fr}
\date{Received date / Accepted date}

  \abstract {
   We analyse the properties of Mg~{\sc ii} absorption
systems detected along the sightlines toward GRBs using a sample of 10 GRB afterglow 
spectra obtained with VLT-UVES over the past six years. The signal-to-noise ratio is
sufficiently high that we can extend previous studies to smaller equivalent widths 
(typically $W_{\rm r}$~$>$~0.3~\AA). Over a pathlength of $\Delta z\sim14$, we detect 9 intervening 
Mg~{\sc ii} systems with $W_{\rm r}$~$>$~1~\AA~ and 9 weaker MgII systems (0.3~$<$~$W_{\rm r}$~$<$~1.0~\AA) when about 4 and 7, respectively, are expected from observations
of QSO sightlines. The number of weak absorbers is similar along
GRB and QSO lines of sight, while the number of strong systems is larger along GRB lines of sight
with a 2$\sigma$ significance.
Using intermediate and low resolution observations reported in the literature, we increase the absorption length for strong systems to $\Delta z$~=~31.5 (about twice the path length of previous studies) 
and find that the number density of strong Mg\,{\sc ii} systems is a factor of
2.1$\pm$0.6 higher (about 3$\sigma$ significance)
toward GRBs as compared to QSOs, about twice smaller however than previously reported.
We divide the sample in three redshift bins and we find that the number density of strong Mg~{\sc ii} is larger in the low redshift bins.
We investigate in detail the properties of strong Mg~{\sc ii} systems observed with UVES, 
deriving an estimate of both the H~{\sc i} column density and the associated 
extinction. Both the estimated dust extinction in strong GRB Mg~{\sc ii} systems and
the equivalent width distribution are 
consistent with what is observed for standard QSO systems. 
We find also that the number density of (sub)-DLAs per unit redshift in the UVES 
sample is probably twice larger than what
is expected from QSO sightlines which confirms the peculiarity of GRB lines of sight.
These results indicate that neither a dust extinction bias nor different beam sizes of the sources are viable explanations for the excess. 
It is still possible that the current sample of GRB lines of sight is biased by a subtle gravitational lensing effect. 
More data and larger samples are needed to test this hypothesis.
}
  \titlerunning{ MgII absorbers along GRB lines of sight observed with UVES}
\maketitle

\section{Introduction}
Thanks to their exceptional brightness, and although fading very rapidly,
Gamma-Ray Burst (GRB) afterglows can be used as powerful extragalactic background sources.
Since GRBs 
can be detected up to very high redshifts \citep{Greiner2008a,Kawai2006,Haislip2006} their afterglow spectra can be used 
to study the properties and evolution of galaxies and the IGM, similarly to what is 
traditionally done using QSO spectra.

Even if the number of available GRB lines of sight (los) is much smaller than those of QSOs, 
it is interesting to compare the two types of lines of sight.
In particular, 
\cite{Prochter2006a} found that the number density of strong (rest equivalent width $W_{\rm r}>1$\,\AA) 
intervening Mg~{\sc ii} absorbers is more than 4 times larger along GRB los than what is expected for 
QSOs over the same path length. This result has been derived from a sample of
14 GRB los and a redshift path of $\Delta z=15.5$, and has been confirmed by
\cite{Sudilovsky2007}. Dust extinction bias for QSO los, gravitational lensing, contamination 
from high-velocity systems local to the GRB and difference of beam sizes are among the possible 
causes of this discrepancy. All these effects can contribute to the observed excess, but
no convincing explanation has been found to date for the amplitude of the excess 
\citep{Prochter2006a,Frank2007,Porciani2007}. 
Similar studies \citep{Sudilovsky2007,Tejos2007} have been performed on the number of C~{\sc iv} 
intervening systems. Their results are in agreement with QSO statistics.

Clearly, further investigation of this excess is required. Since the reports by \cite {Prochter2006a} 
(based on a inhomogeneous mix of spectra from the literature) and the confirmation by \cite{Sudilovsky2007} 
(based on a homogeneous but limited sample of UVES los), several new los have been observed. 
As of June 2008, the number of los with good signal-to-noise ratio observed by UVES has increased to 10. 
We use this sample to investigate the excess of strong Mg~{\sc ii} absorbers and, thanks to the  
high quality of the data, we also extend the search of systems to lower equivalent widths and
derive physical properties of the absorbing systems. In addition, to increase the redshift path over which 
strong Mg~{\sc ii} systems are observed, we consider also a second sample that includes in addition other observations 
gathered from the literature.

We describe the data in Section\,\ref{Data}, identify Mg~{\sc ii}
systems and determine their number density in Section~\ref{Number}. We derive some characteristics 
of strong Mg~{\sc ii} systems 
in Section~\ref{pop}, we estimate their HI content in Section\,\ref{HIstrong} and we study peculiar (sub-)DLAs systems detected along the lines of sight
in Section\,\ref{DLAend}. We summarize and conclude in Section~\ref{Concl}.

\section{Data}
\label{Data}

Our first sample (herafter the {\it UVES sample}) includes ten GRB afterglows 
with available follow-up VLT/UVES\footnote{UVES is described in \citet{Dekker2000}.} high-resolution
optical spectroscopy
as of June 2008: GRB\,021004, GRB\,050730, GRB\,050820A, GRB\,050922C, GRB\,060418, GRB\,060607A, 
GRB\,071031, GRB\,080310, GRB\,080319B and GRB\,080413A.
All GRBs were detected by the \emph{Swift} satellite \citep{Gehrels2004}, with the 
exception of GRB~021004, which was detected by the \emph{High-Energy Transient Explorer (HETE-2)} satellite 
\citep{Ricker2003}.

UVES observations
began on each GRB afterglow with the minimum possible time delay.
Depending on whether the GRB location was immediately observable from Paranal, and whether UVES was observing at the time of the GRB explosion, the afterglows were observed in either Rapid-Response Mode (RRM) or as Target-of-Opportunity (ToO). 
A log of the observations is given in Table 1.

\begin{table}[htp]
\tiny
\begin{minipage}[t]{18cm}
\caption{GRB sample and log of UVES observations}
\begin{tabular}{lccccc }
\hline\hline
GRB & UT\footnote{UT of trigger by the BAT instrument on-board
  \emph{Swift}. Exception: \newline GRB~021004, detected by \emph{WXM} 
  on-board \emph{HETE-2}.} &  
$\delta t$\footnote{Time delay between the satellite trigger and the start
  of the first \\ UVES exposure: normally a series of spectra is taken.} & $t_{\rm total}$\footnote{Total UVES exposure
  time including all instrument setups.} & ESO Program&PI \\
(yymmdd) & \emph{Swift} & (hh:mm) & (h) &ID& \\ 
\hline
021004  & 12:06:13 &  13:31 &  2.0 & 070.A-0599\footnote{\tiny{Also 070.D-0523 (PI: van den Heuvel).}}&Fiore \\
050730  & 19:58:23 &  04:09 & 1.7 & 075.A-0603 &Fiore\\
050820A  & 06:34:53 &  00:33 &  1.7 & 075.A-0385&Vreeswijk \\
050922C & 19:55:50 & 03:33 &  1.7 & 075.A-0603&Fiore\\ 
060418 & 03:06:08 & 00:10&2.6 &077.D-0661&Vreeswijk\\
060607A  & 05:12:13 &  00:08 & 3.3 & 077.D-0661& Vreeswijk \\
071031  & 01:06:36 & 00:09 &  2.6 & 080.D-0526 &Vreeswijk\\
080310  & 08:37:58 & 00:13 &  1.3 & 080.D-0526& Vreeswijk\\
080319B &06:12:49&00:09  & 2.1 &080.D-0526\footnote{\tiny{Also 080.A-0398 (PI: Fiore).}}&Vreeswijk\\
080413A &02:54:19 &03:42 & 2.3&081.A-0856&Vreeswijk\\

\hline
\end{tabular} 
\end{minipage}

\end{table}

The observations were performed with a 1.0\arcsec\ wide slit and 2x2 binning, providing a resolving power of $R\approx48\,000$ (FWHM~$\sim$~7~km s$^{-1}$) 
for a $\approx$2~km s$^{-1}$\ pixel size\footnote{Though the minimum guaranteed resolving power of
UVES in this mode is 43\,000, we find that in some cases a higher resolution, up to
$\approx$50\,000, is achieved in practice, due to variations in seeing conditions.}.
The UVES data were reduced with a customized version of the {\sc MIDAS}
reduction pipeline \citep{Ballester2006}.
The individual scientific exposures were co-added by
weighting them according to the inverse of the variance as a function of wavelength and rebinned in the heliocentric rest frame.

Although the UVES sample has a smaller number of los compared to the sample
used by \cite{Prochter2006a} (10 instead of 14 los), the redshift path
of the two samples is similar ($\Delta z$~=~13.9 and 15.5 for the UVES
and Prochter's samples respectively) because of the larger wavelength coverage of the UVES spectra. 
The UVES sample has 6 new los (050922C, 060607A, 071031, 080310,
080319B and 080413A) corresponding to $\Delta z$~=~9.4 that are not in the Prochter's sample. 
Therefore more than two third of the redshift path is new. In addition, the UVES 
sample is homogeneous (similar resolution, same instrument, similar signal-to-noise ratio) and more than doubles 
the sample used by \cite{Sudilovsky2007}, which is included in our sample, for the Mg~{\sc ii} statistics.

The second sample we consider (the {\it overall sample}) is formed 
by adding observations from the literature (see Table \ref{taball}) to the {\it UVES sample}
(see Section~3.2).
The sample gathers observations of 26 GRBs for $\Delta z$~=~31.55. It 
therefore doubles the statistics of \cite{Prochter2006a}. 

In the following we use solar abundances from \cite{Grevesse2007}.

\section{Number density of Mg~{\sc ii} absorbers}
\label{Number}

\subsection{The UVES sample}
\label{uves}

For each line of sight we searched by eye the spectrum for
Mg~{\sc ii} absorbers outside the Lyman-$\alpha$ forest considering all 
Mg~{\sc ii} components within 500 km s$^{-1}$ as a single system. 
Table~\ref{riasMgII} summarizes the results. 
Columns 1 to 8 give, respectively, the name of the GRB, its redshift, the
redshift paths along the line of sight for $W_{\rm r, lim}$~$>$~0.3~ and 1~\AA~ (see Eq.~\ref{wlim}),
the redshift of the Mg~{\sc ii} absorber, the rest equivalent width of
the Mg~{\sc ii}$\lambda 2796$ transition 
and the velocity relative to the GRB redshift.
The last column gives comments on peculiar systems if need be. 
When a Mg~{\sc ii}$\lambda 2796$ line is blended either with a sky feature or an absorption
from another system, we fit simultaneously Voigt profile components to the 
Mg~{\sc ii} doublet and the contaminating absorption and derive characteristics of
the Mg~{\sc ii}$\lambda 2796$ absorption from the fit.
The sky spectrum at the position of a feature is obtained from other UVES
spectra in which the metal absorption is not present. The upper limits include also the contaminating absorption. 

\begin{table*}[htp]
\caption{Characteristics of the Mg~{\sc ii} absorbers in the UVES sample.}

\centering
{ \small
\begin{tabular}{lccccccl}
\\
\hline
\hline
\bf{GRB}  & $z_{\rm GRB}$  &  $\Delta z $  &$ \Delta z$ & $z_{\rm abs}$ & $W_{\rm r}(\lambda2796)$ & v$_{\rm ej}$  &Remarks   \\ 
          &                &  W$_{\rm r}>$0.3\AA   &      W$_{\rm r}>$1\AA              &             &  (\AA) &(km/s)  & \\ 
\hline 

021004&2.3295 &1.754&1.756& 0.5550 &  $0.248\pm0.025$ & $\sim192000$&blended with AlII1670 at $z=1.6026$ \\
&& & &  1.3800 &  $1.637\pm0.020$  &$\sim97000$ &\\
&& & & 1.6026&  $1.407\pm0.024$ &$\sim72000$& \\
\hline 

050730 &3.9687&1.278&1.298& 1.7732 &  $0.927\pm0.030$  &$\sim157000$& \\
 & & & &  2.2531 &  $<0.783\, (0.650)$& $\sim120000$&sky contamination subtracted\\  
\hline 

050820A & 2.6147 &1.843& 1.845&0.6896 & $0.089\pm0.007$  & $\sim192000$&\\
&&&&0.6915 &  $2.874\pm0.007$ &$\sim192000$&\\
&& & & 1.4288 &  $1.323\pm0.023$  & $\sim113000$&\\
 & & & &  1.6204 &  $0.277\pm0.024$ &  $\sim93000$&\\ 
 & & & &  2.3598 &  $<0.424\,(0.306)$ &  $\sim22000$&contribution by FeII2600  at $z=2.6147$\\ 
 &&&&&&& subtracted\\
\hline 
 
050922C & 2.1996 &1.679&1.682 &0.6369 &  $0.179\pm0.018$ &$\sim175000$& \\
&& &  & 1.1076 &  $0.677\pm0.029$ & $\sim118000$&\\
 & &&& 1.5670 &  $<0.102\, (0.08)$ & $\sim62000$&sky contamination subtracted \\ 
\hline 

060418 & 1.4900 &1.242 &1.265 &0.6026 &  $1.293\pm0.010$  &$\sim124000$ &\\
&& && 0.6559 &  $1.033\pm0.006$  & $\sim116000$&\\
 & & &  &1.1070 &  $1.844\pm0.014$  & $\sim50000$&\\ 
\hline 
 
060607A & 3.0748 & 1.710&1.713& 1.5103 &  $0.204\pm0.011$ &$\sim135000$& \\
&& &&  1.8033 &  $1.854\pm0.006$ & $\sim107000$&\\
 &&& & 2.2783 & $0.343\pm0.058$  &$\sim64000$ &\\

\hline 
  
071031 & 2.6922&1.727 &1.789 & 1.0743&  $0.330\pm0.016$ &$\sim156000$&\\
&& & &   1.6419&  $0.692\pm0.014$ &$\sim97000$&\\ 
 & & & & 1.9520 & $0.804\pm0.016$ &$\sim66000$ &\\

\hline 
  
080310 & 2.4272& 1.841&1.841 &1.1788&  $0.047\pm0.024 $&$\sim127000$& \\
&& &&    1.6711&  $0.421\pm0.012$ &$\sim73000$&\\  

\hline 
  
080319B & 0.9378&0.790 &0.790 &  0.5308&$ 0.614\pm0.001$ &$\sim69000$ &\\
&& &&  0.5662&  $0.083\pm0.003$ &$\sim63000$&\\
&& & &  0.7154&  $1.482\pm0.001$ &$\sim36000$&\\ 
 & & & &  0.7608& $0.108\pm0.002$  & $\sim29000$&\\ 
  \hline 
 080413A & 2.4346&1.583 &1.650 & 2.1210&$ 0.768\pm0.259$ &$\sim29000$ &\\

  \hline  
  \hline
\end{tabular}
}
\label{riasMgII}
\end{table*}%

The rest equivalent width detection limit (at any given statistical level)
is calculated at each redshift along the spectrum
using the following equation \citep{Tytler1994}:

\begin{equation}
W_{\rm r,lim} \simeq {U M_{\rm L}^{0.5} \over SNR} {\rm \Delta} \lambda\,\,  \rm({\AA}),
\label{wlim}
\end{equation}

\noindent
where $M_{\rm L}$ is the number of pixels the line is expected to cover, $U$ is the number of 
rms-intervals (or $\sigma$) defining the statistical significance of the detection limit, $SNR$ is the signal-to-noise ratio
at the corresponding wavelength and $\Delta\lambda$ is the FWHM of the resolution element. 
We will apply this detection limit to the Mg~{\sc ii}$\lambda$2796 transition.
Using $M_{\rm L}=5$ and $U=6$ (so that each transition of the Mg~{\sc ii} doublet 
is detected at more than 3~$\sigma$), it can be seen that 
for a typical signal-to-noise ratio of SNR~$\sim$~10 and a typical FWHM of 
$\Delta\lambda$~=~0.13~\AA~ ($R$~=~43000 and $\lambda$~=~5600~\AA), 
the formula gives a detection limit of $\sim$0.17~\AA~
which is much smaller than the equivalent width limit we will use
in our statistical studies below (1 and 0.3~\AA).

We then can compute the redshift path over which a line of a given equivalent width 
would be detected in our data.  
\cite{Prochter2006a} limited their analysis to the redshift range $0.359 - 2$ 
and compared their results to those of the SDSS QSO survey reported by \cite{Prochter2006}. However,
the SDSS QSO MgII survey extends to $z_{\rm max}$~$\sim$~2.3 \citep{Nestor2005} so that there is no reason 
to limit our analysis to $z=2$ and we will use  the same redshift limits as 
Nestor et al. (2005), $ z_{\rm start}=0.366$ and $z_{\rm max}$~=~2.27 instead.
As in all QSO surveys and following Prochter et al. (2006a), we will exclude along each line-of-sight 
the redshift range within an ejection velocity of 3000 km s$^{-1}$ from the GRB redshift.
Table~\ref{numball} lists the mean redshift, $\langle z_{\rm abs}\rangle$, the total redshift paths obtained considering these redshift limits, and 
the number of systems detected in our sample over these redshift paths for different $W_{\rm r}$ 
limits or ranges: $W_{\rm r}>0.3$\,\AA \,(\textit{'all systems'}),
$W_{\rm r}>1.0$\,\AA\,(\textit{'strong systems'}) and $0.3\le W_{\rm r}\le 1.0$\,\AA \,(\textit{'weak systems'}). 
The total redshift paths for the \textit{all} and \textit{strong} samples are 
$\Delta z=13.79$ and $13.94$, respectively, for an observed number, $N^{\rm MgII}_{\rm obs}$, of 18 and 9 systems detected, corresponding
to redshift number densities of $\partial n/\partial z=1.31\pm0.31$, 0.65$\pm$0.22 
for \textit{all} and \textit{strong} Mg~{\sc ii} systems, respectively.

\begin{table}[htp]
\caption{Number of Mg~{\sc ii} systems and redshift paths} 
{ 
\centering
\tiny
\begin{tabular}{lccc}
\hline
\hline
\\
$W_{\rm r}(\lambda2796)$ & $>$0.3\,\AA & $>$1\,\AA & $>0.3$ and $<$1 \AA \\ 
\\
\hline 
\\
$\langle z_{\rm abs}\rangle$&1.34&1.11&1.57\\
\\
\hline 
\\Redshift path    & 13.79 & 13.94 & 13.79 \\[1ex]
\hline
\\
 $N^{\rm MgII}_{\rm obs}$ (UVES sample)& 18   & 9   & 9   \\
\\
\hline
\\
$N^{\rm MgII}_{\rm exp}$&$11.98\,(\pm3.46 $)&$4.83\,(\pm2.20$)&$7.21\,(\pm2.68 $)\\[1ex]
(Nestor et al., 2005)  &&&\\
\hline
\\
$N^{\rm MgII}_{\rm exp}$&$$&$4.00\,(\pm2.00$)&$$\\[1ex]
(Prochter et al., 2006)  &&&\\
\hline
\hline
\end{tabular}
}
\label{numball}
\end{table}%

\begin{figure}[htp]
\begin{center}
\includegraphics[width=0.7\linewidth]{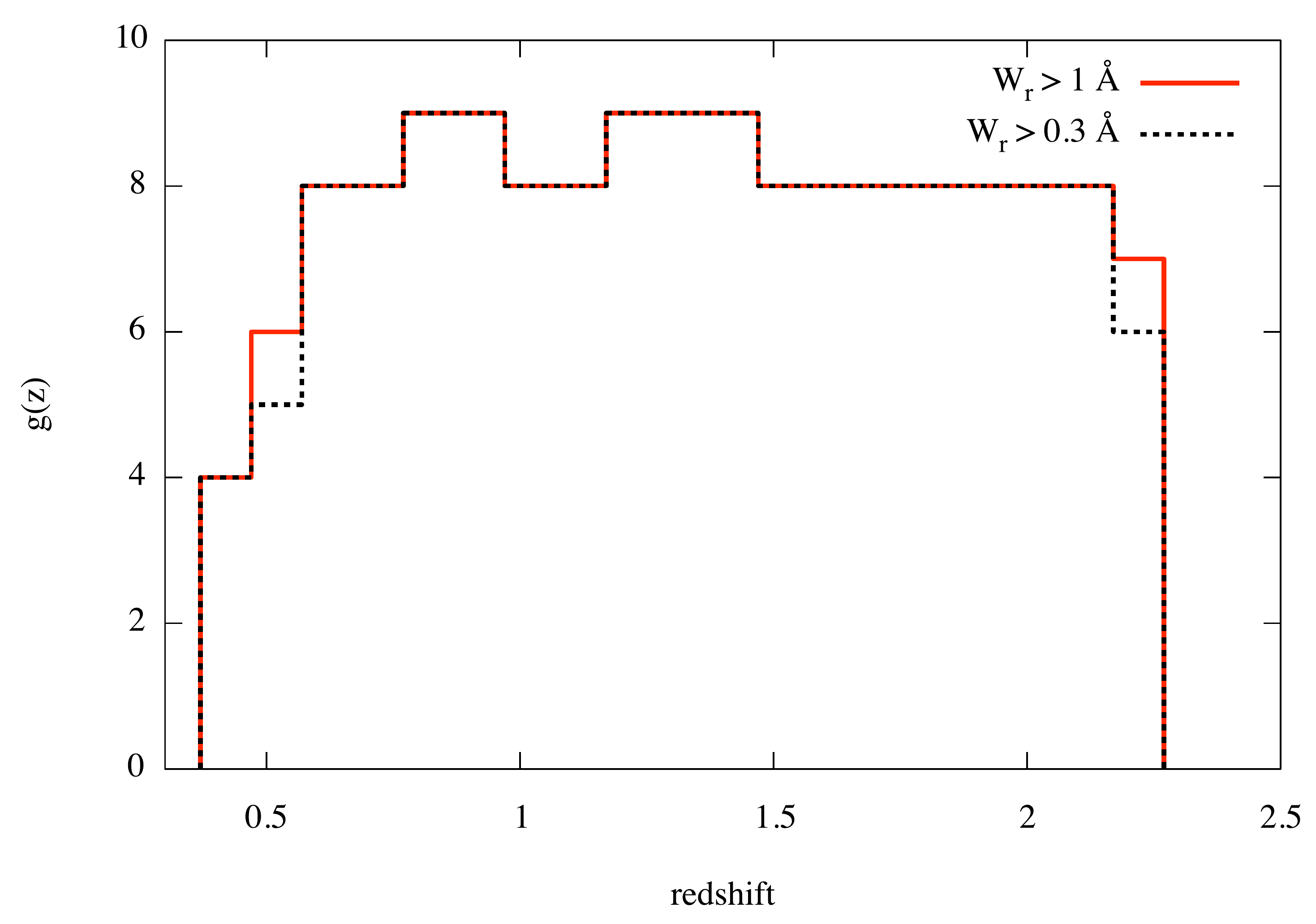}
\caption{Redshift path density $g(z)$ of the {\it UVES sample} 
for $W_{\rm r}>0.3$\,\AA \,(black dashed line) and $W_{\rm r}>1.0$\,\AA\, (red line).}
\label{gz}
\end{center}
\end{figure}

\begin{figure}[htp]
\begin{center}
\includegraphics[width=0.7\linewidth]{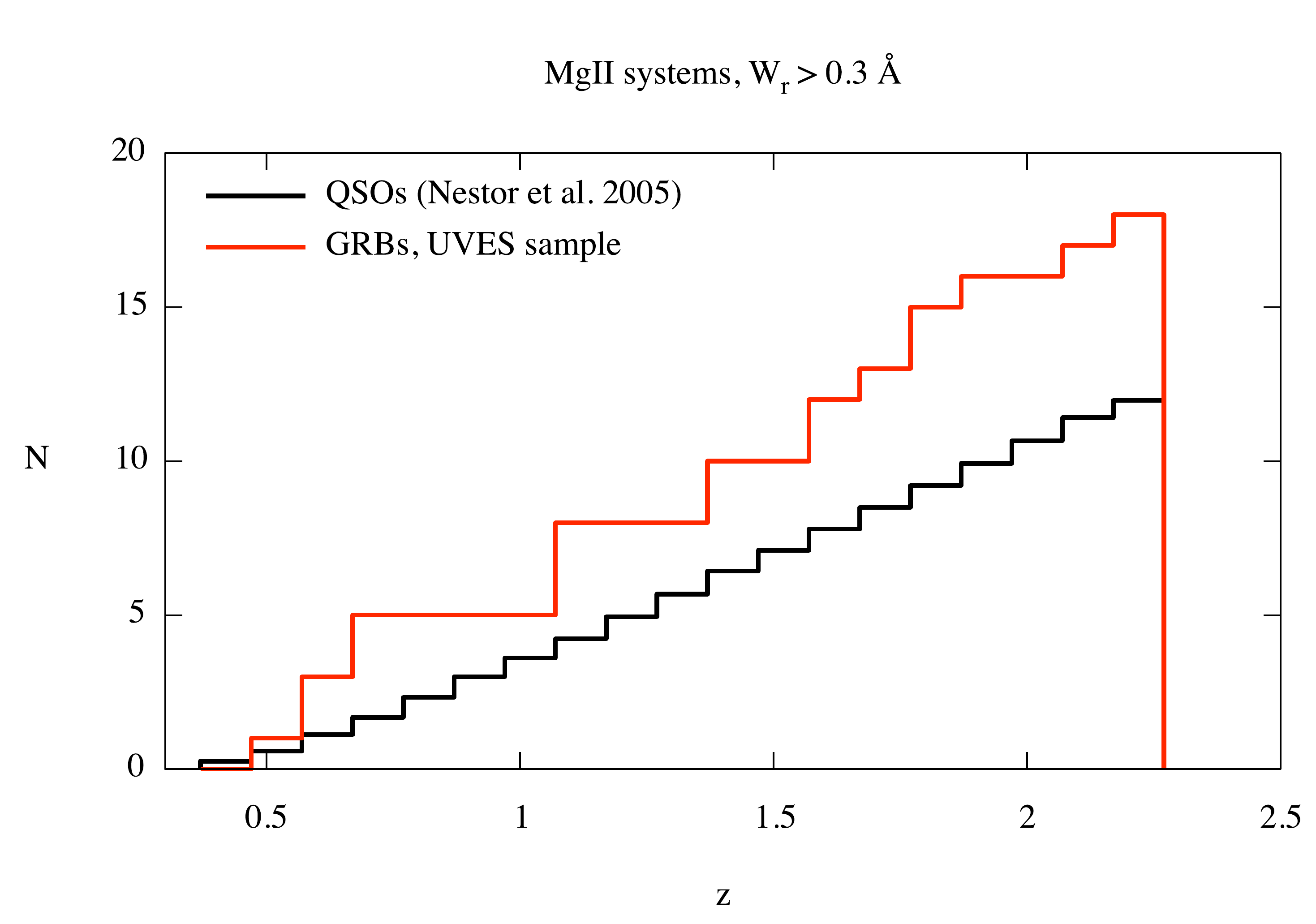}
\hfill
\includegraphics[width=0.7\linewidth]{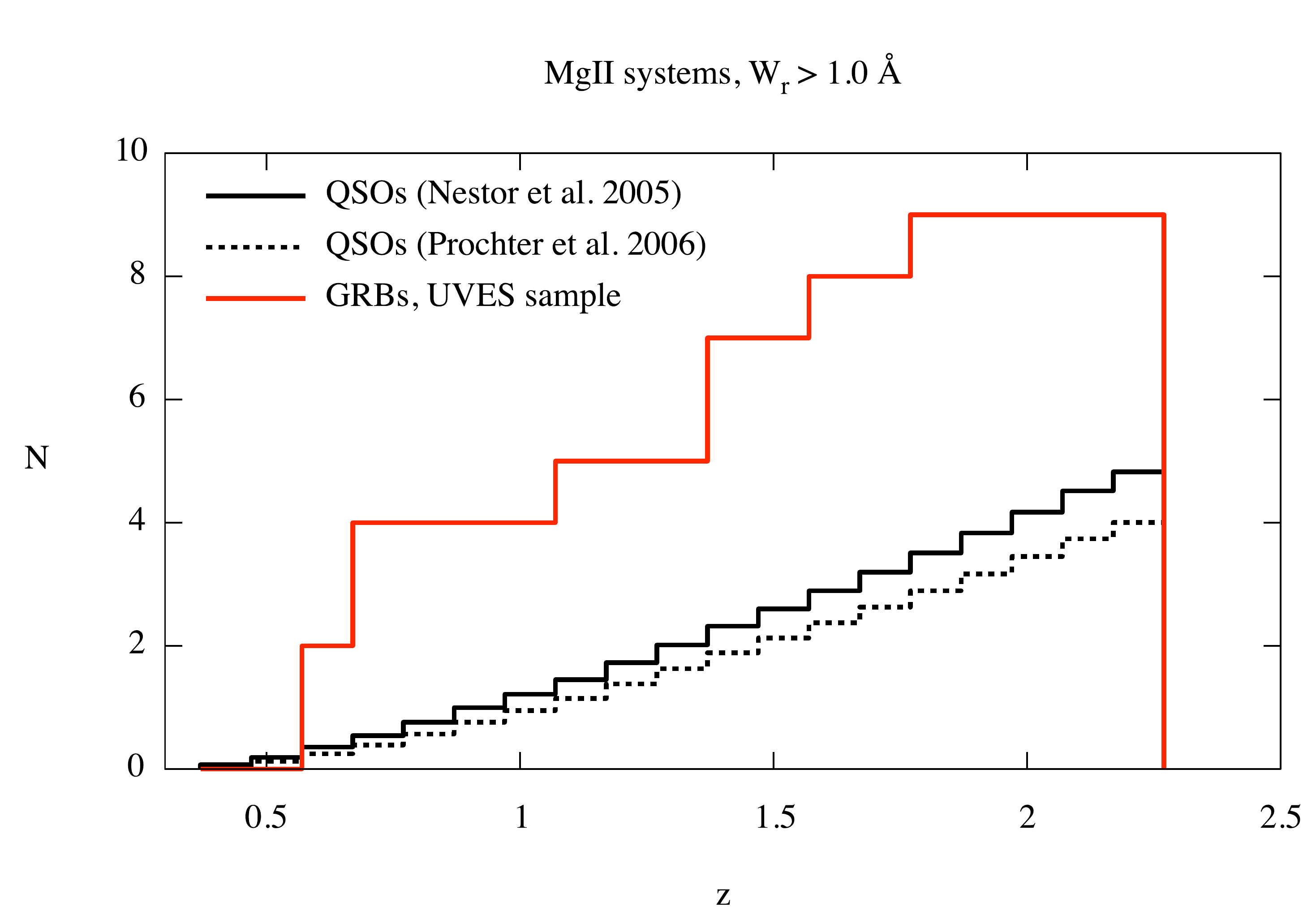}
\hfill
\includegraphics[width=0.7\linewidth]{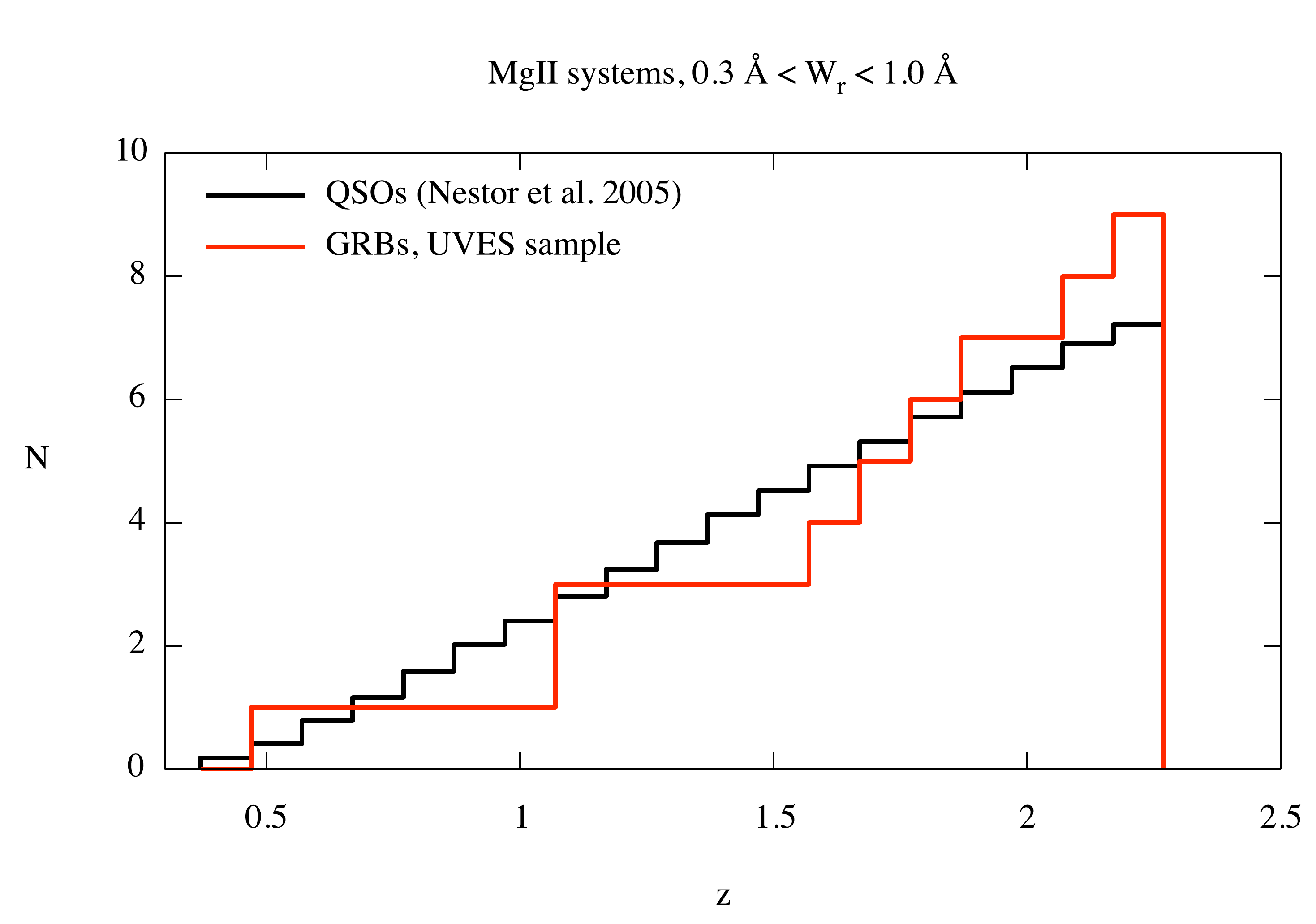}
\hfill
\caption{Comparison between the cumulative distribution of Mg~{\sc ii} systems detected along the UVES 
GRB los (red) and the one expected along QSO los following Eq.\,2 (black solid) or Eq.\,3 (black dashed), for:
\textit{all} (top panel), \textit{strong} (middle panel) and \textit{weak} (bottom panel) systems.}
\label{all}
\end{center}
\end{figure}

In Table~\ref{numball} we also report the number of Mg~{\sc ii} absorbers that would be expected along
lines of sight toward QSOs over the same redshift path, $N^{\rm MgII}_{\rm exp}$.
To calculate these numbers, we determine the redshift path density, $g(z)$, of the UVES GRB sample 
(see Fig. \ref{gz})
and combine it with the density of QSO absorbers per unit redshift as observed by \cite{Nestor2005} and 
\cite{Prochter2006} (for the strong systems only) in the SDSS survey:

\begin{equation}
N^{\rm MgII}_{\rm exp}=\int_{z_{\rm start}}^{z_{\rm end}} \! g(z) \frac{\partial n}{\partial z}  \, dz.
\end{equation}

Both \cite{Nestor2005} and \cite{Prochter2006} showed that the traditional parametrization of 
$\partial n/\partial z$ as a simple powerlaw $n_0(1+z)^{\gamma}$ does not provide a good fit to the SDSS data.
Therefore we will use their empirical fits to the redshift number density.
\cite{Nestor2005} give

\begin{equation}
\frac{\partial n}{\partial z}  = \int\frac {dn} {dW_{\rm r}}(z)dW = \int\frac{n^*(z)} {W^*(z)} e^{-W_r/W^*}dW, 
\end{equation}

\noindent
where both $n^*$ and $W^*$ vary with redshift as power laws:
$n^*=1.001\pm 0.132 (1+z)^{0.226\pm0.170}$, $W^*=0.443\pm 0.032 (1+z)^{0.634\pm0.097} $.
\cite{Prochter2006a} derive for the strong MgII systems:

\begin{equation}
\partial n/\partial z  = -0.026+0.374z-0.145z^2+0.026z^3 
\end{equation}

The results of the calculations for the different $W_{\rm r}$ ranges 
($W_{\rm r}>0.3$\,\AA \,, $W_{\rm r}>1.0$\,\AA\, and $0.3\le W_{\rm r} \le 1.0$\,\AA) are shown in Fig. \ref{all} and 
reported in Table~\ref{numball}. Errors are assumed to be poissonian and are scaled as
$\sqrt{N^{\rm MgII}_{\rm exp}}$.

As a verification, we note 
from Fig.\,13 of \cite{Nestor2005} that the number density, $\partial n/\partial z $, 
of $W_{\rm r}>0.3$\,\AA\, systems is almost independent of the redshift. Multiplying 
the $\partial n/\partial z=0.783 $ value reported there by the UVES  
\textit{'all systems'} $\Delta z = 13.79$ we obtain an expected total 
number of 9.78, in agreement with the value reported in 
Tab.~\ref{numball} for systems along QSO los (i.e. 10.54). The same test performed using the number density values found by \cite{Steidel1992} give consistent results. 

Fig.~\ref{all} and Table\,3 show that 
the excess of \textit{strong} Mg~{\sc ii} systems along GRB los compared to QSO los is 
significant at more than $2\sigma$ (slightly less than $2\sigma$ for the \textit{strong} Mg~{\sc ii} systems if the \citealt{Nestor2005} function is used), but it is more than a factor of $\sim 2$ 
lower than what is  found by \cite{Prochter2006a}. The number of weak systems is consistent 
within $\sim1\sigma$ to that expected for QSO los. 

\subsection{The overall sample} 

\begin{table*}[htp]
\caption{GRB los available from the literature.}
{
\begin{center}
\small
\begin{tabular}{cccccccc}
\hline
\hline
GRB &  $z_{\rm GRB}$  & $z_{\rm start}$ & $z_{\rm end}$ & $z_{\rm abs}$ & $W_{\rm r}(2796)$ & 
$\Delta v_{\rm ej}$   & Reference $^\dagger$\\
    &                 &  &   &   &  (\AA) & (km/s) & \\
\hline
991216& 1.022 & 0.366 & 1.002 &0.770$^a$  & $4.0 \pm 0.8^b$  & $\sim40000$ & 3\\
& &&  & 0.803  & $6.1 \pm 0.7^b$ &  $\sim34000$&  \\
000926 & 2.038 & 0.616 & 2.008 &  &  &  & 1\\
010222 & 1.477 & 0.430 & 1.452  & 0.927  & $1.00 \pm 0.14$ &$\sim 74000$& 2\\
       & &&  & 1.156  & $2.49 \pm 0.08$ & $\sim41000 $&  \\
011211 & 2.142 & 0.366 & 1.932 &  &  &   & 3 \\
020405 & 0.695 & 0.366 & 0.678 & 0.472 & $1.1 \pm 0.3$ &$\sim 42000$ & 4 \\
020813 & 1.255 & 0.366 & 1.232  & 1.224 & $1.67 \pm 0.02$ & $\sim4000$  & 5 \\
030226 & 1.986 & 0.366 & 1.956 & & & & 6 \\
030323 & 3.372 & 0.824 & 1.646  & & & & 7 \\
030328& 1.522 & 0.366& 1.497  & & & & 14 \\
030429 & 2.66 & 0.620 & 1.861  & 0.8418    & $3.3\pm0.4^b$ & $\sim179000$ & 15 \\
050505 & 4.275 & 1.414 & 2.27  & 1.695    & $1.98$ & $\sim176000$  & 8 \\
& &&  & 2.265 & $1.74 $ & $\sim134000 $ & \\
050908 & 3.35 & 0.814 & 2.27  & 1.548  & $1.336 \pm 0.107$ & $\sim147000$  & 9\\
051111 & 1.55 & 0.488 & 1.524  & 1.190  & $1.599 \pm 0.007$ & $\sim45000$  & 9,13\\
060206 & 4.048 & 1.210 & 2.27  & 2.26  & $1.60$ & $\sim123000$ & 10\\
060526 & 3.221 & 0.836 & 2.27  &  & &   & 11\\
071003 & 1.604 & 0.366 & 1.578  & 0.372  & $2.48 \pm 0.20$ & $\sim170000$  & 12\\
\hline
\hline
\end{tabular}

\smallskip
 $^\dagger$ 1: \cite{Castro2003}; 2: \cite{Mirabal2002}; 3: \cite{Vreeswijk2006a}; 5: \cite{Barth2003}; 4: \cite{Masetti2003}; \\ 6: \cite{Klose2004}; 7: \cite{Vreeswijk2004}; 8: \cite{Berger2006}; 9: \cite{Prochter2006a}; 10: \cite{Chen2008}; \\ 11: \cite{Thoene2008}; 12: \cite{Perley2008}; 13: \cite{Hao2007}; 14: \cite{Maiorano2006}; 15: \cite{Jakobsson2004}.

$^a$The very low resolution of the GRB\,991216 spectrum makes the $z$=0.77 absorber identification uncertain.
$^b$The $W_{\rm r}$ values refer to the total equivalent width of the MgII doublet.

\end{center}
}
\label{taball}
\end{table*}

\begin{figure}[htp]
\begin{center}
\includegraphics[width=0.8\linewidth]{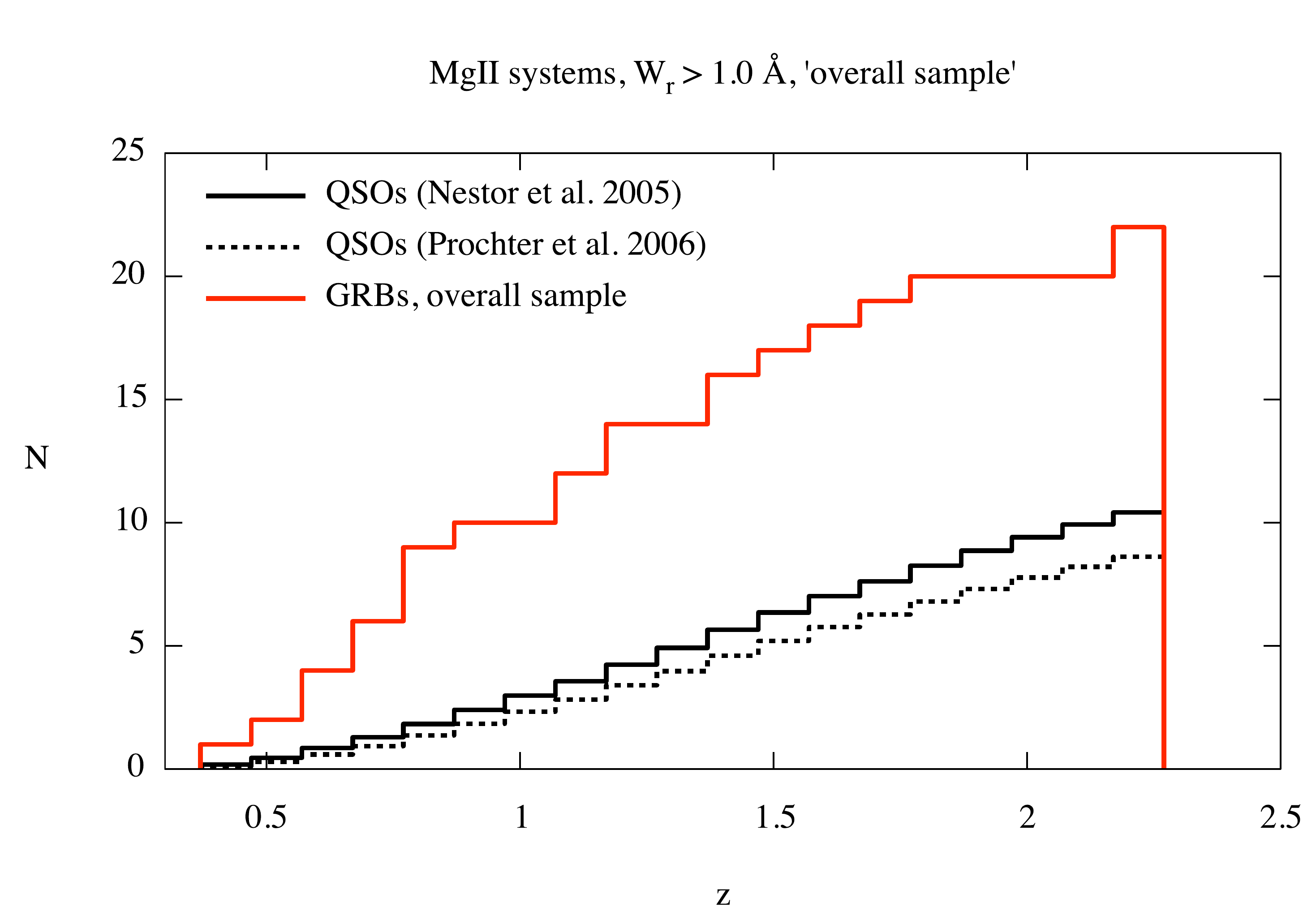}
\caption{Comparison between the cumulative distribution of strong Mg~{\sc ii} systems in the 
overall sample of GRB los (red) and the one expected along QSO los following Eq.\,2 (black solid) 
or Eq.\,3 (black dashed).}
\label{overall}
\end{center}
\end{figure}

In order to increase the statistics of strong Mg~{\sc ii} systems,
we added to the UVES sample both high and low resolution GRB afterglow spectra published in the literature. 
The resulting sample is composed of 26 los (see Table~\ref{taball} for details on los that are not
part of the UVES sample). Since this sample includes many 
low resolution spectra, we study the statistics of strong ($W_{\rm r}$~$>$~1~\AA) Mg~{\sc ii} absorbers only, 
using the same redshift limits as in the previous Section. Three lines of sight of this sample 
(GRB\,991216, GRB\,000926 and GRB\,030429) were not used by
\cite{Prochter2006a} because the low spectral resolution of these spectra 
does not allow to resolve the Mg~{\sc ii} doublet. However, such a low resolution does not prevent the
detection of strong systems, although the doublet is blended. 
In addition, the total equivalent width of the doublets detected along these 
los is larger than 3~\AA~so that we are confident that $W_{\rm r,2796}$ is larger than 1\,\AA. 
In any case, as detection of Mg~{\sc ii} systems is more difficult
at lower resolution, including these lines of sight could only underestimate their number density.

The total number of strong Mg~{\sc ii} systems is $N=22$ and the redshift path is $\Delta z=31.55$.
This leads to a number density $\partial n/\partial z = 0.70\pm0.15$. We use the $g(z)$ function 
of this enlarged sample to compute the total number of strong systems expected for a similar QSO sample 
following the same method as used in Sect.~\ref{uves}. We find $N^{\rm MgII}_{\rm exp}=10.41\pm3.23$ and $N^{\rm MgII}_{\rm exp}=8.62\pm2.94$, 
using Eq. 3 and 4, respectively, that is $2.1\pm0.6$ and $2.6\pm0.8$ times 
less than the number found along GRB los (see Fig. \ref{overall}). 

The excess of strong Mg~{\sc ii} systems along GRB lines of sight for this enlarged sample is 
confirmed at a $\sim 3\sigma$ statistical significance. The excess found is higher than for the 
UVES sample but still a factor of $\sim2$ lower than what was previously reported by \cite{Prochter2006a} 
from a sample with a smaller redshift path. The number densities resulting from the two studies are different by no more than 2$\sigma$. For consistency we also performed our analysis considering only the smaller sample used by  \cite{Prochter2006a}. In this case, the results obtained are similar to those found by these authors. The redshift path of our overall sample is twice as large 
as that used by \cite{Prochter2006a}, therefore the factor of 2 excess we find in this study has a higher statistical significance.

\subsection{Number density evolution}

\begin{figure}[htp]
\begin{center}
\includegraphics[width=0.6\linewidth]{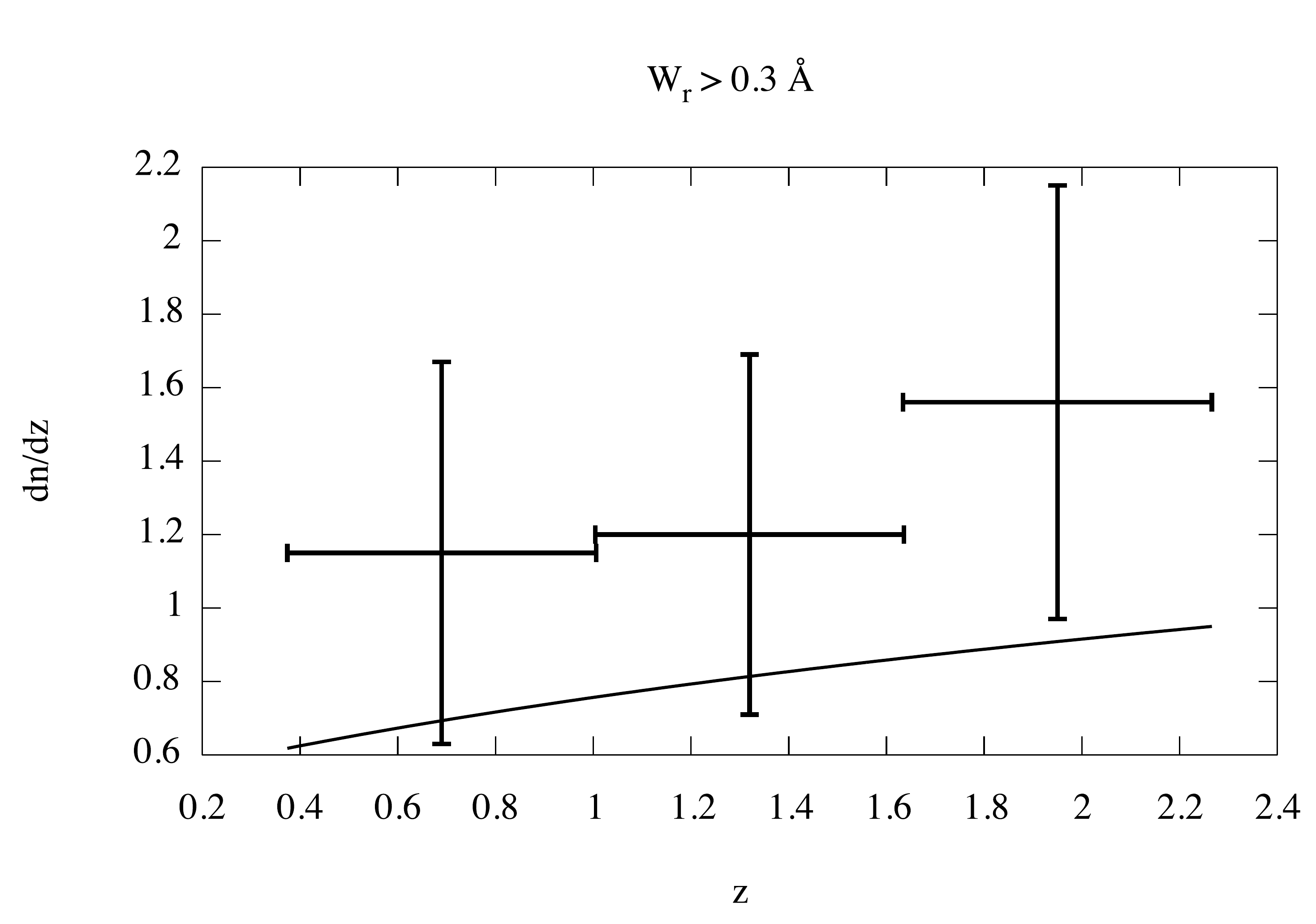}
\hfill
\includegraphics[width=0.6\linewidth]{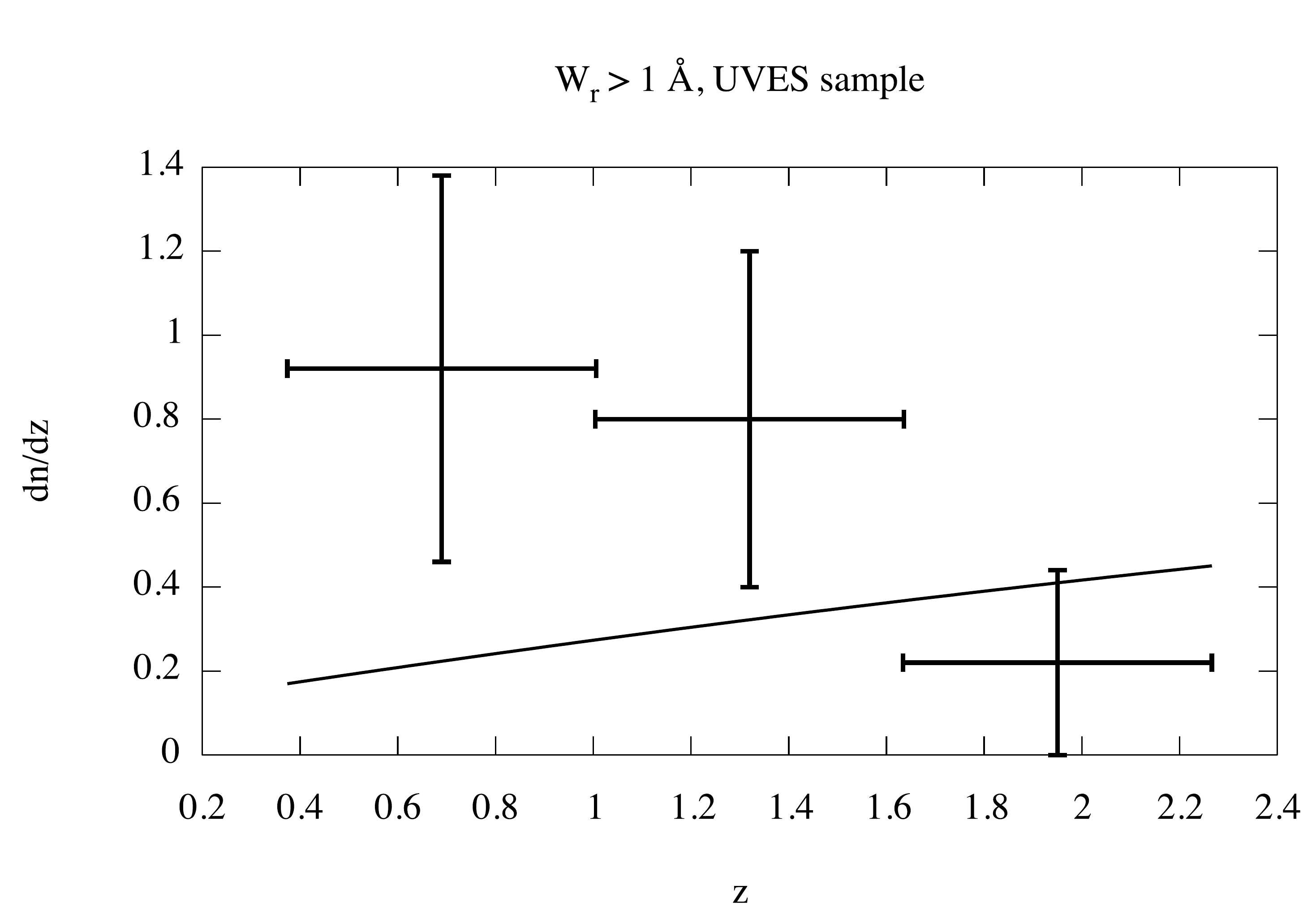}
\hfill
\includegraphics[width=0.6\linewidth]{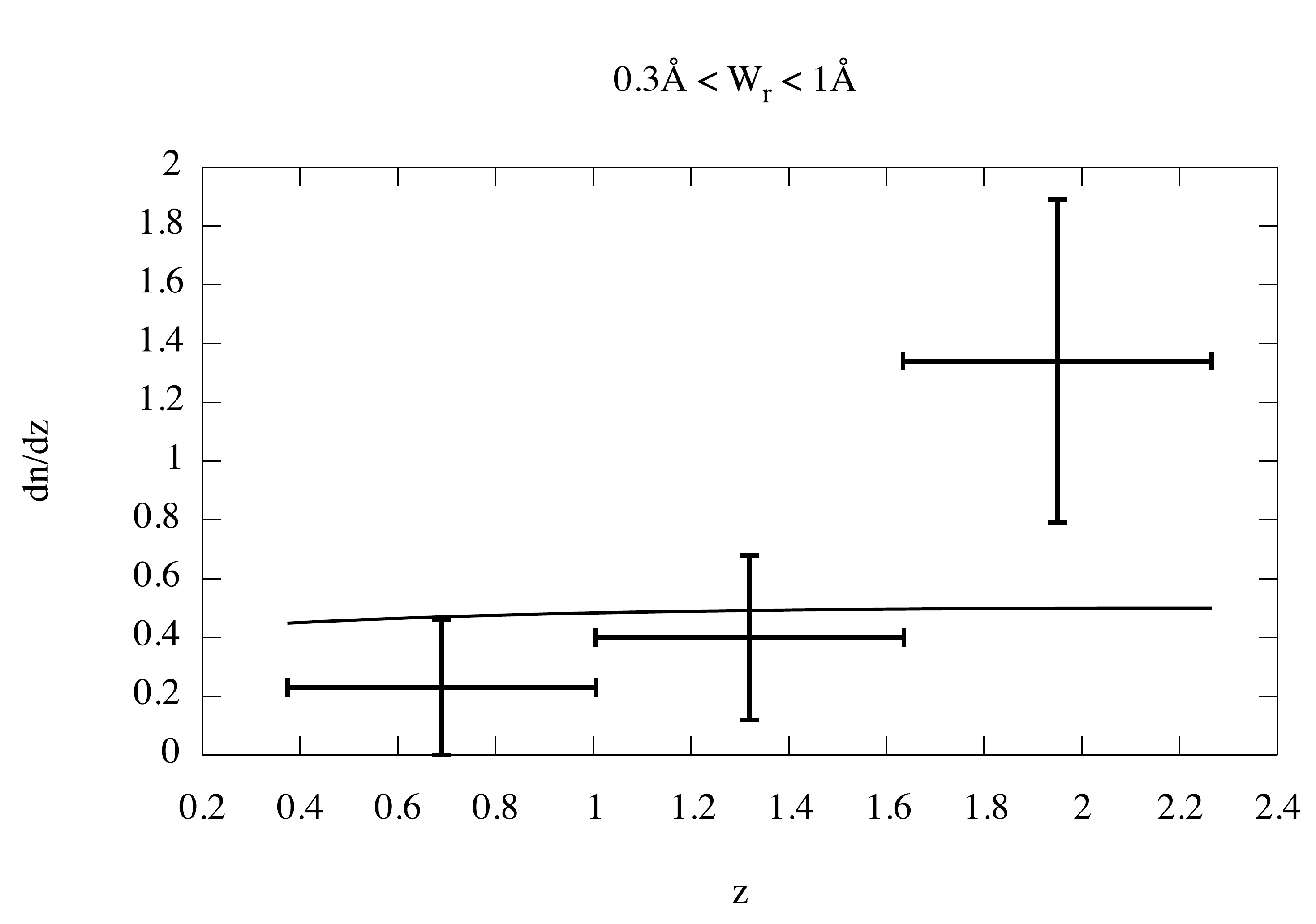}
\hfill
\includegraphics[width=0.6\linewidth]{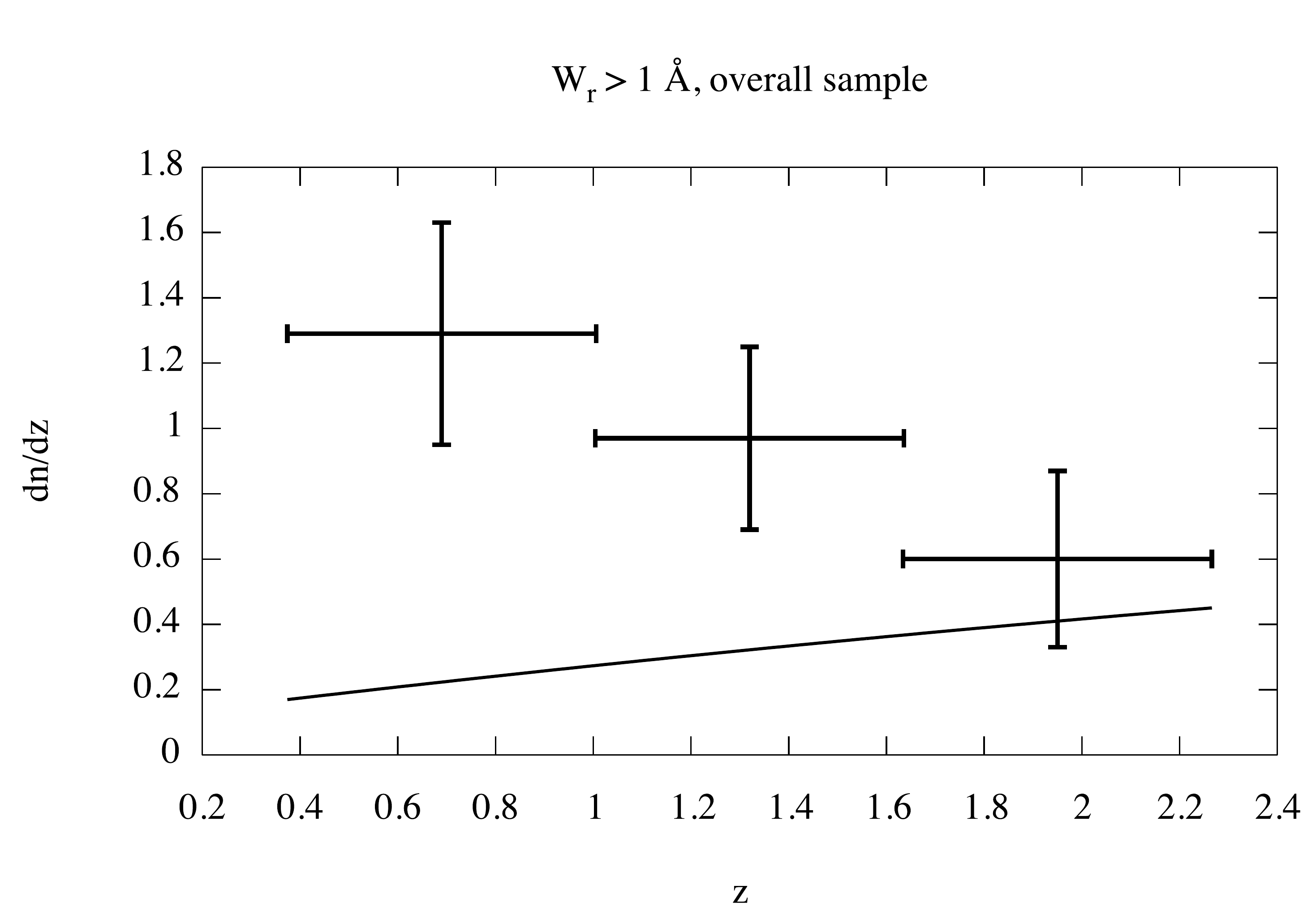}
\caption{Number density evolution of Mg~{\sc ii} systems detected along 
GRB los for: 
first panel: \textit{all}, second panel: \textit{strong}, third panel: \textit{weak} systems of 
the {\it UVES sample}; bottom panel: \textit{strong} systems of the {\it overall sample}.
The solid line represents the evolution of the MgII number density derived from the Mg~{\sc ii} survey in the 
SDSS QSO by \cite{Nestor2005} (see Eq.\,3). }
\label{dndz}
\end{center}
\end{figure}

We divided both the UVES and the overall sample in three redshift bins and calculated 
$\partial n/\partial z$ for each bin. Fig. \ref{dndz} shows the number of systems per redshift bin. 
While the total number of systems and the number of weak systems 
have a comparable redshift evolution in GRB and QSO lines of sight, the
strong systems happen to have a different evolution in GRB los.
The excess of {\it strong} systems in GRBs is particularly pronounced at low redshift, up to $z\sim1.6$.
We performed a KS test for each of the three cases to assess 
the similarity of the redshift distribution of Mg~{\sc ii} systems along GRB and QSO los. 
There is a $90.7\%$ and $23.5\%$ chance that the \textit{weak} and \textit{all} Mg~{\sc ii} 
absorber samples in QSOs and GRBs are drawn from the same population. The probability for 
the \textit {strong} systems is $20.1\%$ for the UVES sample, but it decreases to $\sim2\%$ when 
considering the overall sample. 

This apparent excess of strong systems in the low-redshift bin could indicate that some amplification bias 
due to lensing is at work. Indeed the effect of lensing should be larger in case the deflecting mass is at 
smaller redshift. The lensing optical depth for GRBs at redshift $z_{\rm GRB}>>1$ and  $z_{\rm GRB}\le1$ 
is maximal for a lens at respectively $z_{\rm l}\sim0.7$ and $z_{\rm GRB}/2$. 
However, we find that only 47\% of GRBs with strong foreground absorbers in our sample have at least one strong Mg~{\sc ii}
system located in 
the range within which the optical depth decreases to about half its maximum value \citep{Sudilovsky2007,Turner1980,Smette1997}.
This indicates that if amplification by lensing is the correct explanation, the effect
must be weak and subtle. \cite{Porciani2007} used the optical afterglow luminosities reported by 
\cite{Nardini2006} to show that GRB afterglows with more than one absorbers are brigther than 
others by a factor 1.7. However this correlation has not been detected by \cite{Sudilovsky2007} when using
the afterglow B-band absolute magnitudes obtained by \cite{Kann2006,Kann2007}. 

More data and larger samples are obviously needed to conclude on this issue.

\section{The population of strong Mg~{\sc ii} systems}
\label{pop}

Both the results on the UVES homogeneous sample and those obtained using the overall sample
confirm, although to a smaller extent, the excess of strong Mg~{\sc ii} absorbers along GRB los 
first reported by \cite{Prochter2006a}. 
To understand the reason of the discrepancy in the number density of ``strong'' systems it is 
therefore important to study in more details these systems and to derive their physical 
characteristics. This is possible using the UVES data. 
The main question we would like to address here is whether there is any reason to believe
that GRB and QSO strong absorbers are not drawn from the same population.

\subsection{Equivalent width distribution}

\begin{figure}[htp]
\begin{center}
\includegraphics[width=0.95\linewidth]{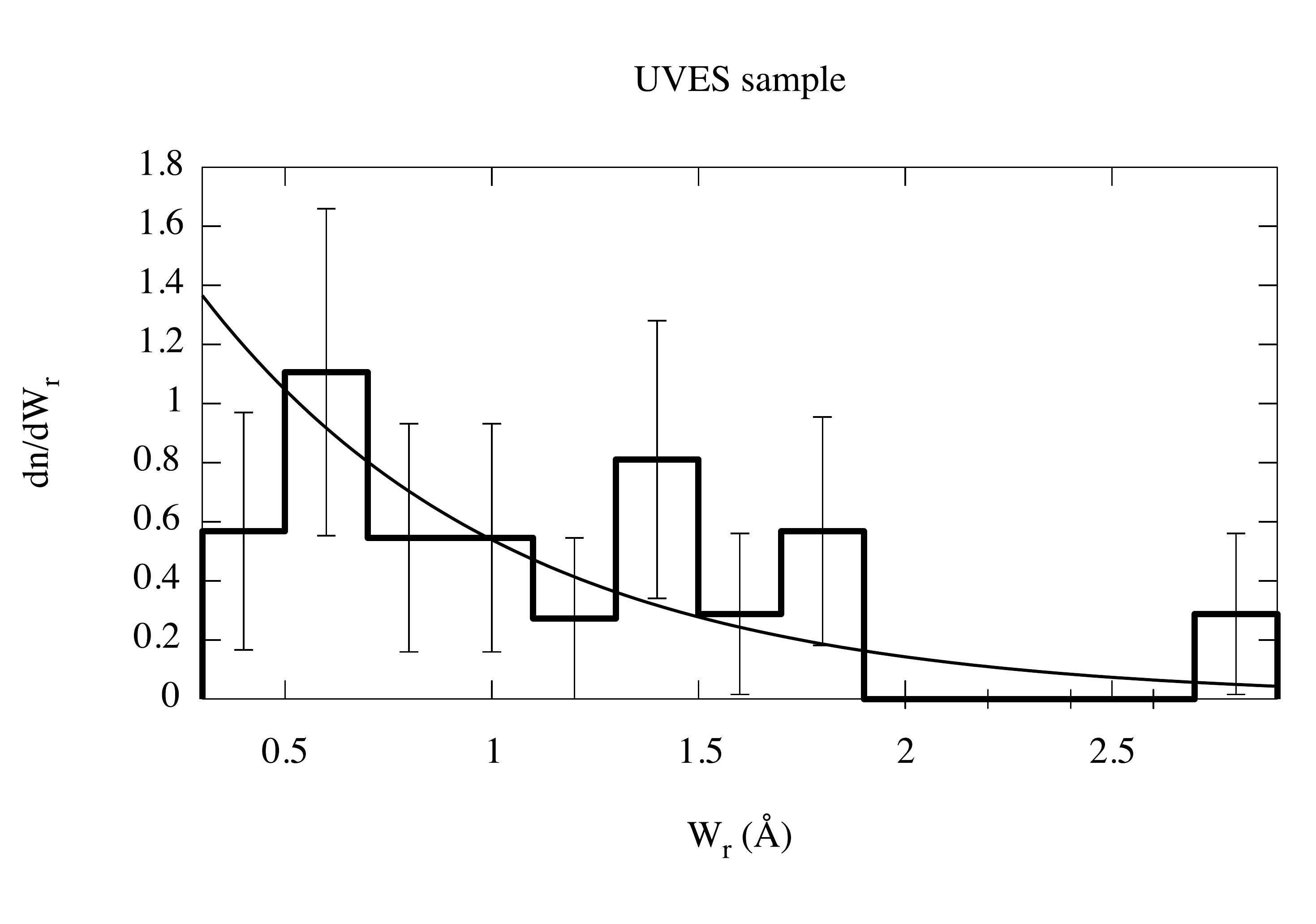}
\hfill
\caption{$W_{\rm r}$ distribution of 
the Mg~{\sc ii} systems with $W_{\rm r}$~$>$~0.3~\AA~ detected in the UVES sample. 
The solid curve 
represents the corresponding distribution for QSO MgII absorbers \citep{Nestor2005}. Both distributions have been normalized such that their underlying areas are equal to 1.
}
\label{wdist}
\end{center}
\end{figure}

We show in Fig.~\ref{wdist} the comparison between the normalized $W_{\rm r}$ distribution of 
all Mg~{\sc ii} systems with $W_{\rm r}$~$>$~0.3~\AA~ detected in the {\it UVES sample} and the one 
reported by \cite{Nestor2005} for the Mg~{\sc ii} systems along the QSO los in the SDSS survey.

The KS tests give a $27.1\%$ 
chance that the GRB and QSO distributions
are drawn from the same population for the {\it UVES sample}.
This result extends to lower $W_{\rm r}$ values the
conclusion by \cite{Porciani2007} that the two
distributions are similar, arguing against the idea that the
excess of Mg~{\sc ii} systems could be related to the internal structure of
the intervening clouds.

\subsection{The velocity spread of strong Mg~{\sc ii} systems}

\begin{figure}[htp]
\begin{center}
\includegraphics[width=0.3\linewidth]{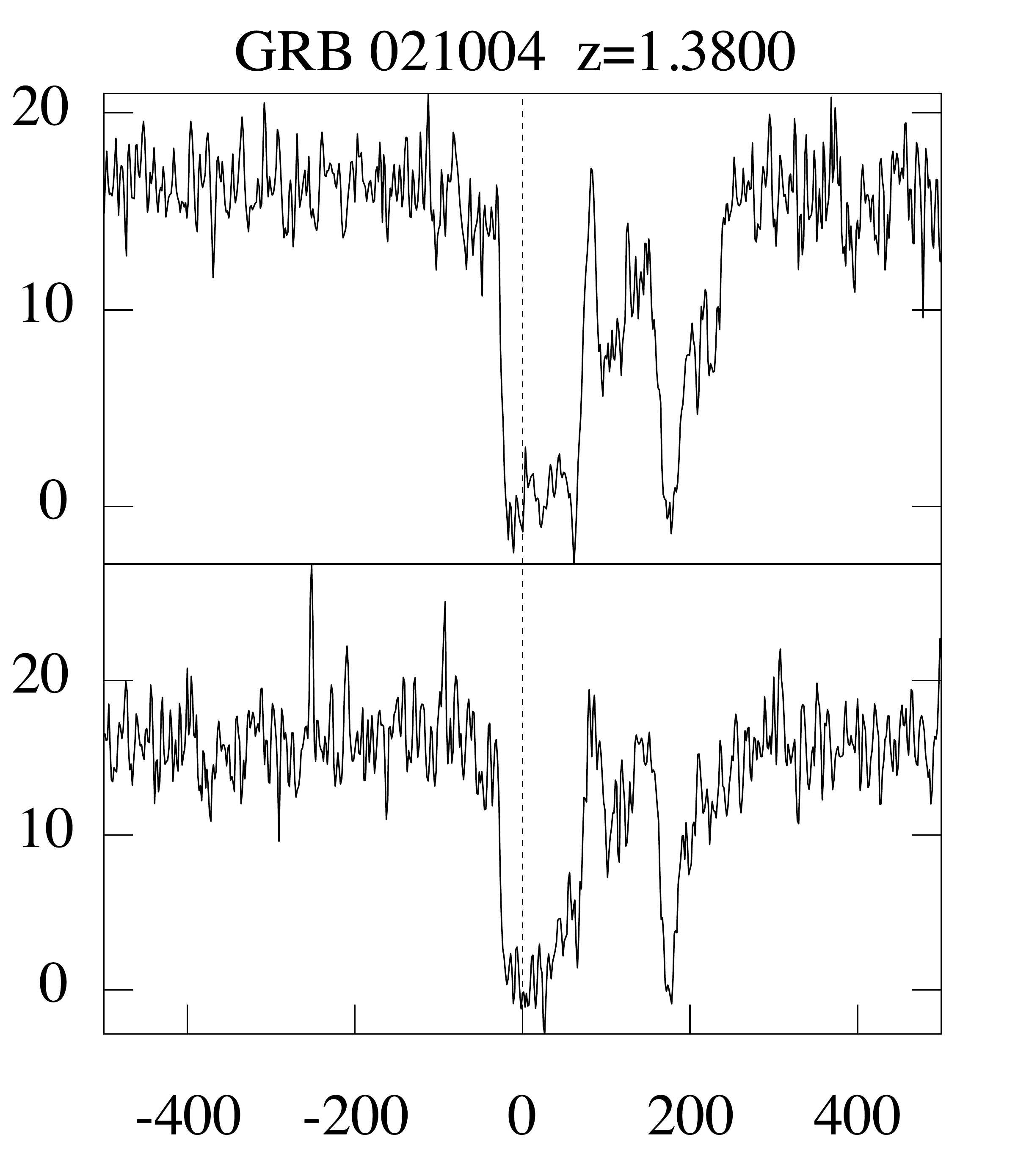}
\includegraphics[width=0.3\linewidth]{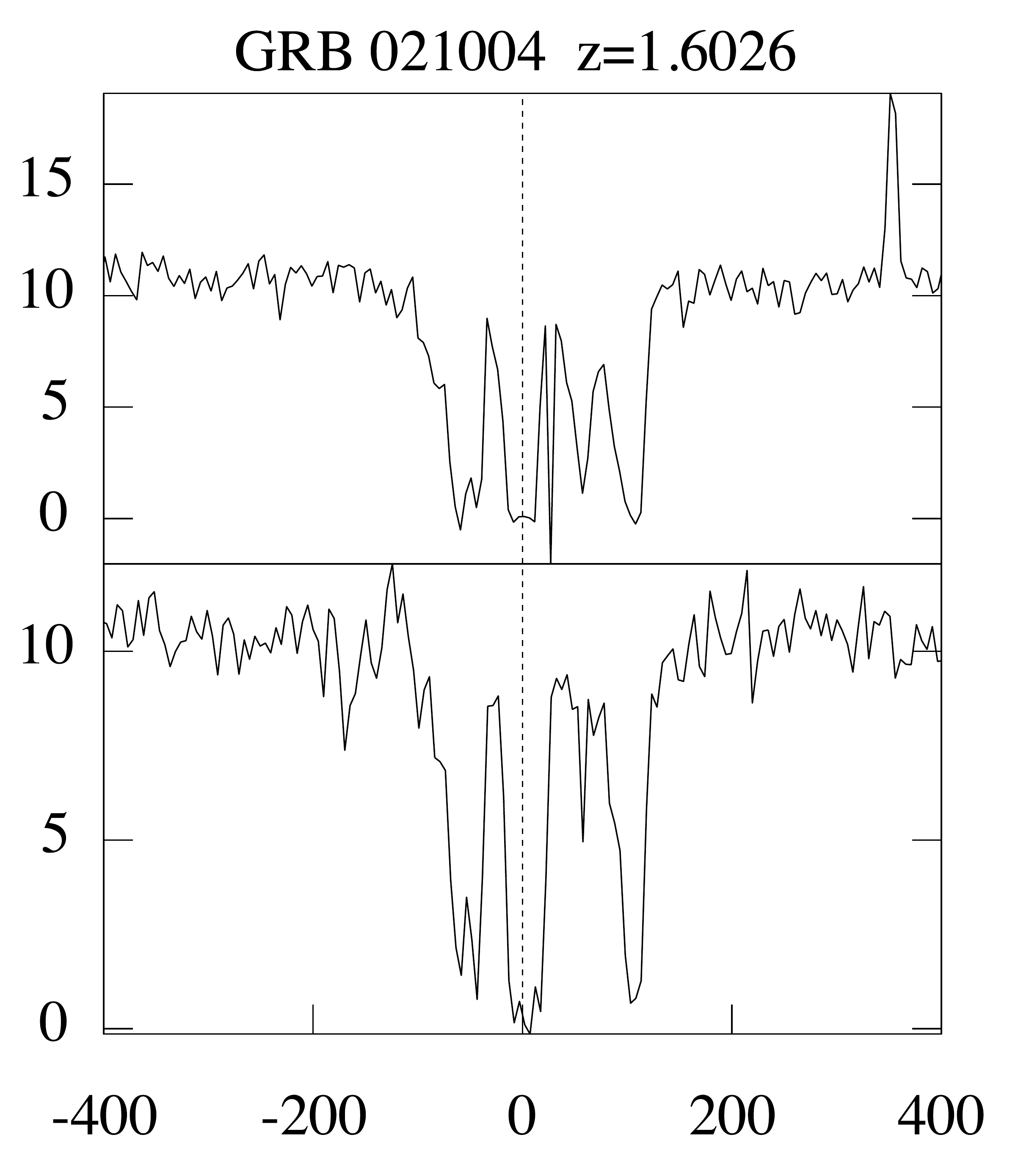}
\includegraphics[width=0.3\linewidth]{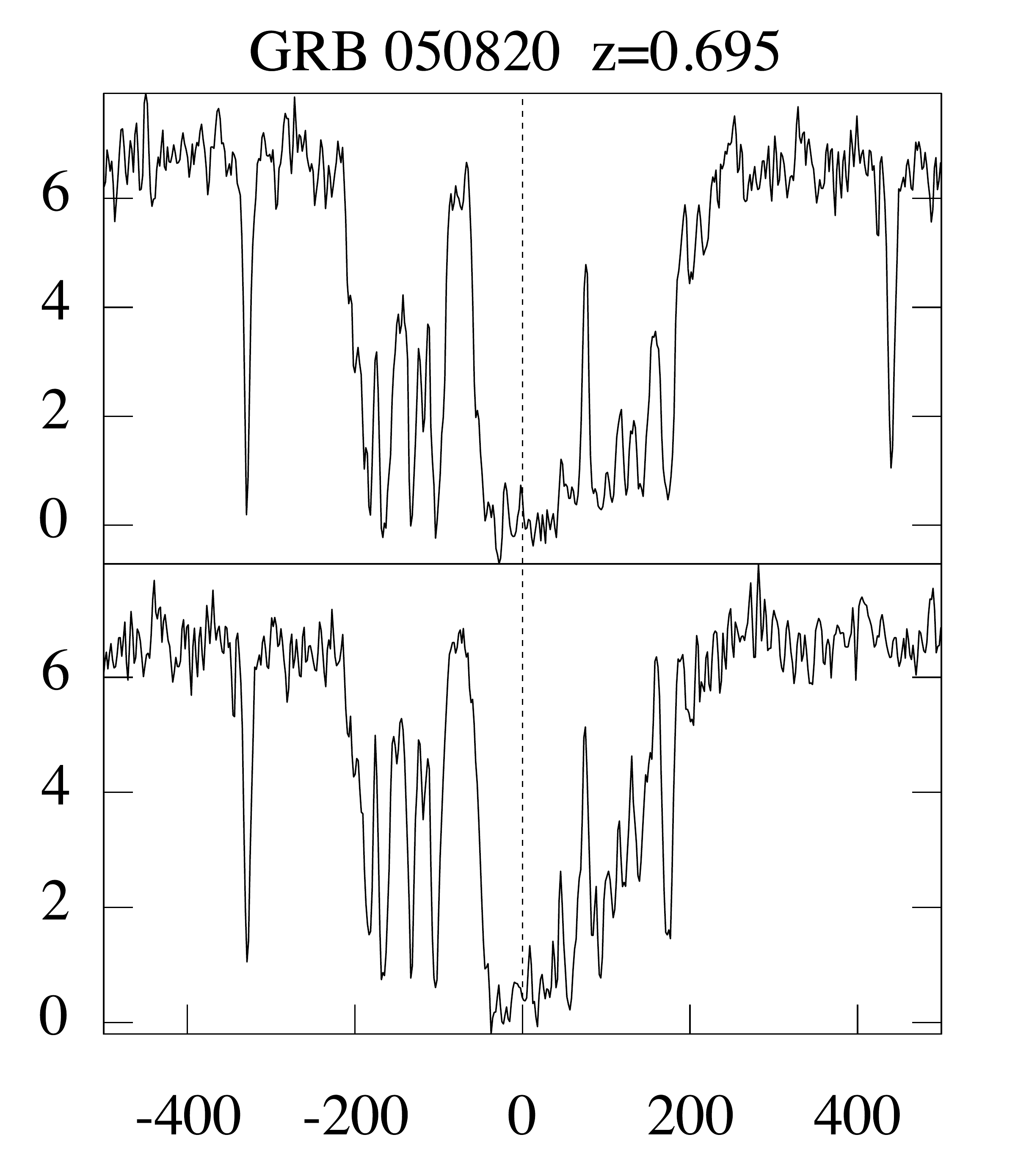}
\includegraphics[width=0.3\linewidth]{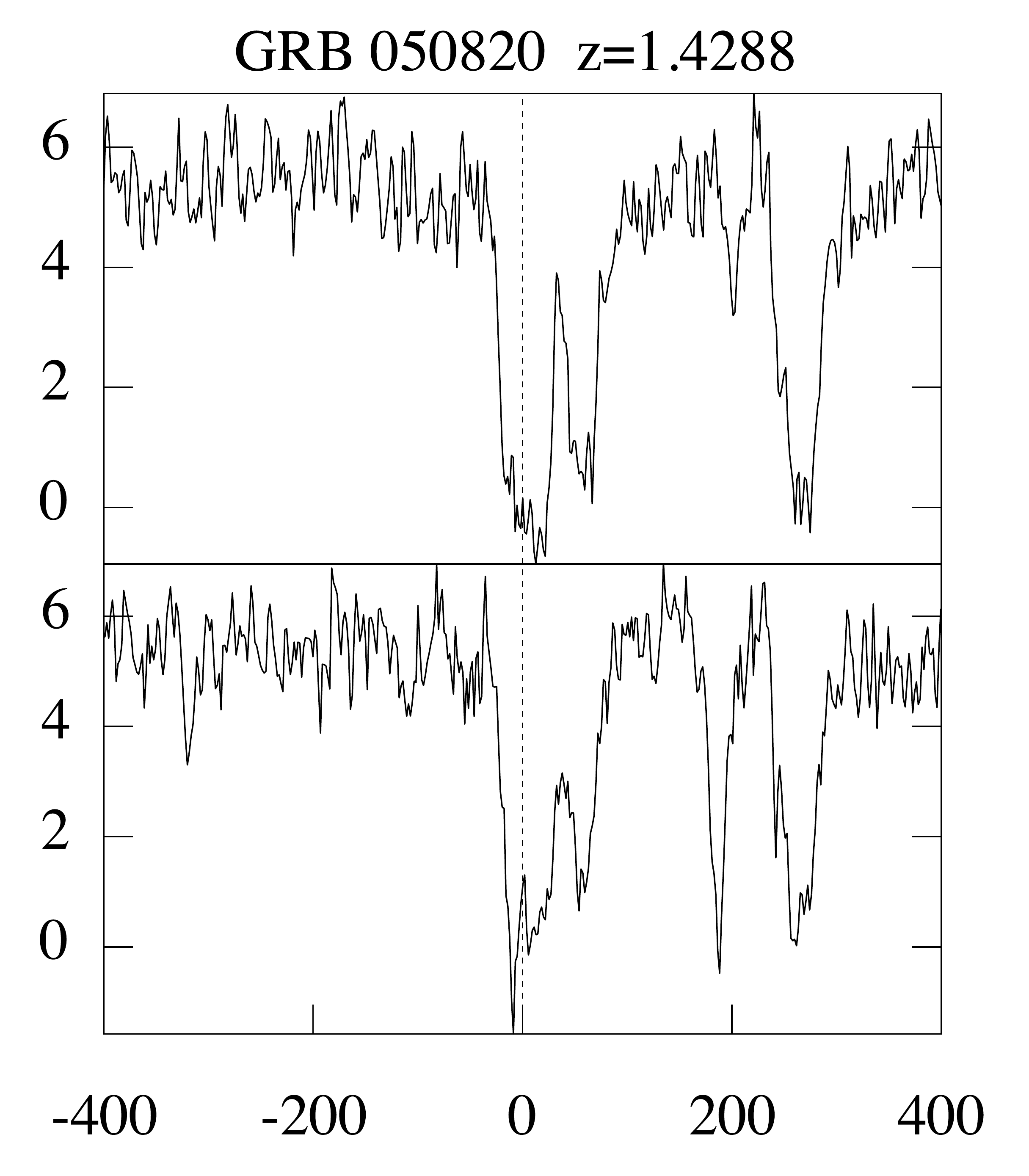}
\includegraphics[width=0.3\linewidth]{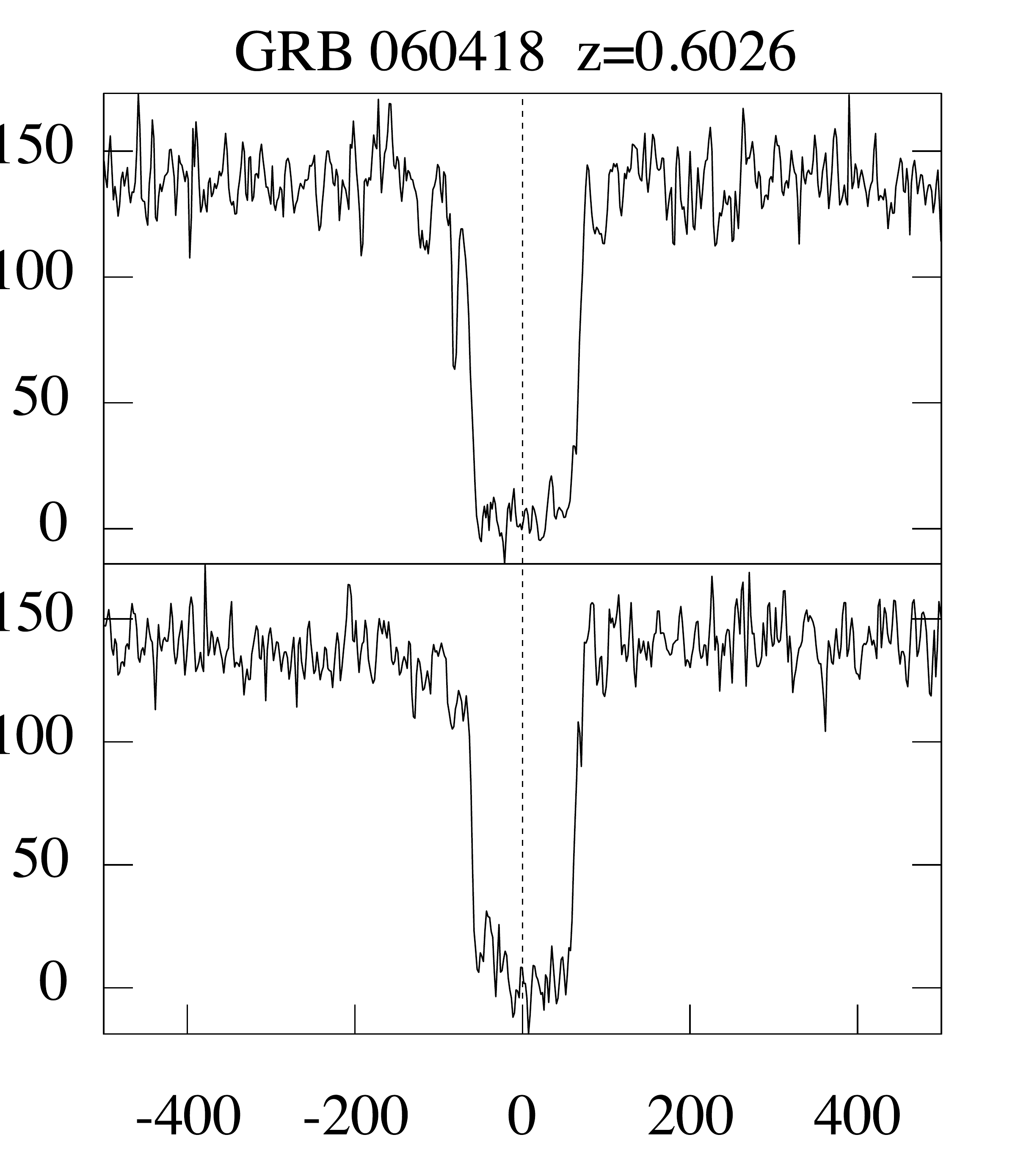}
\includegraphics[width=0.3\linewidth]{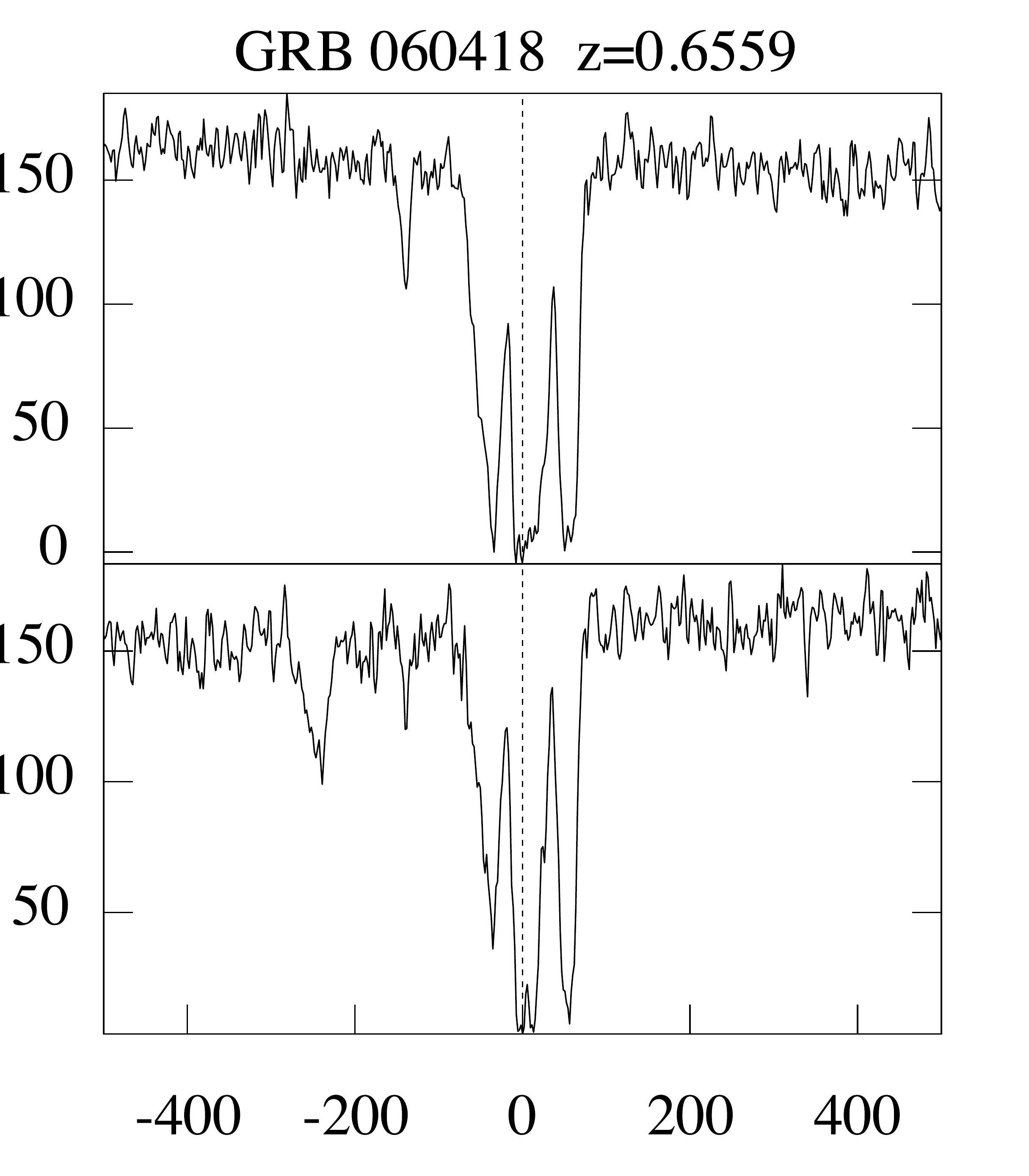}
\includegraphics[width=0.3\linewidth]{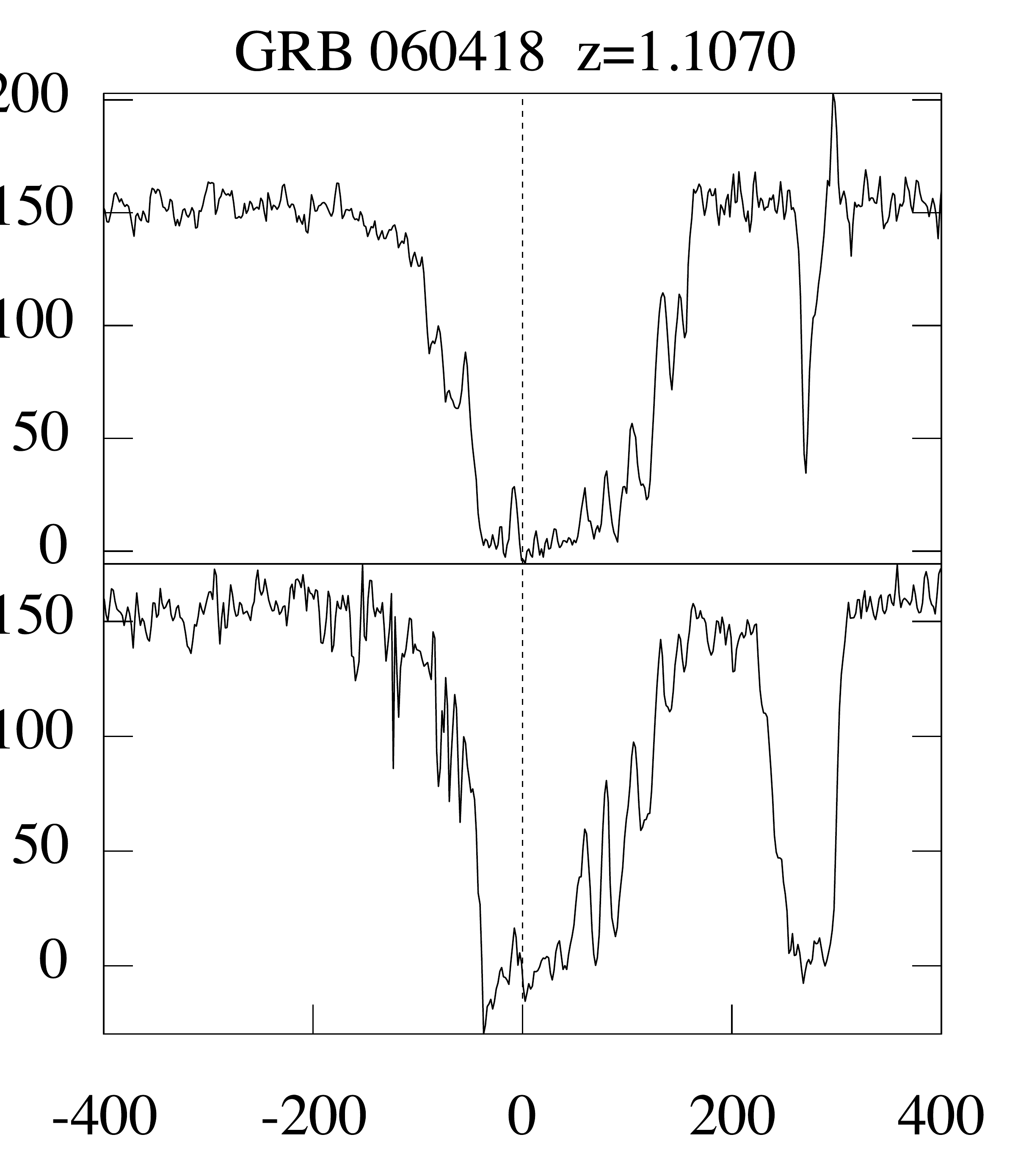}
\includegraphics[width=0.3\linewidth]{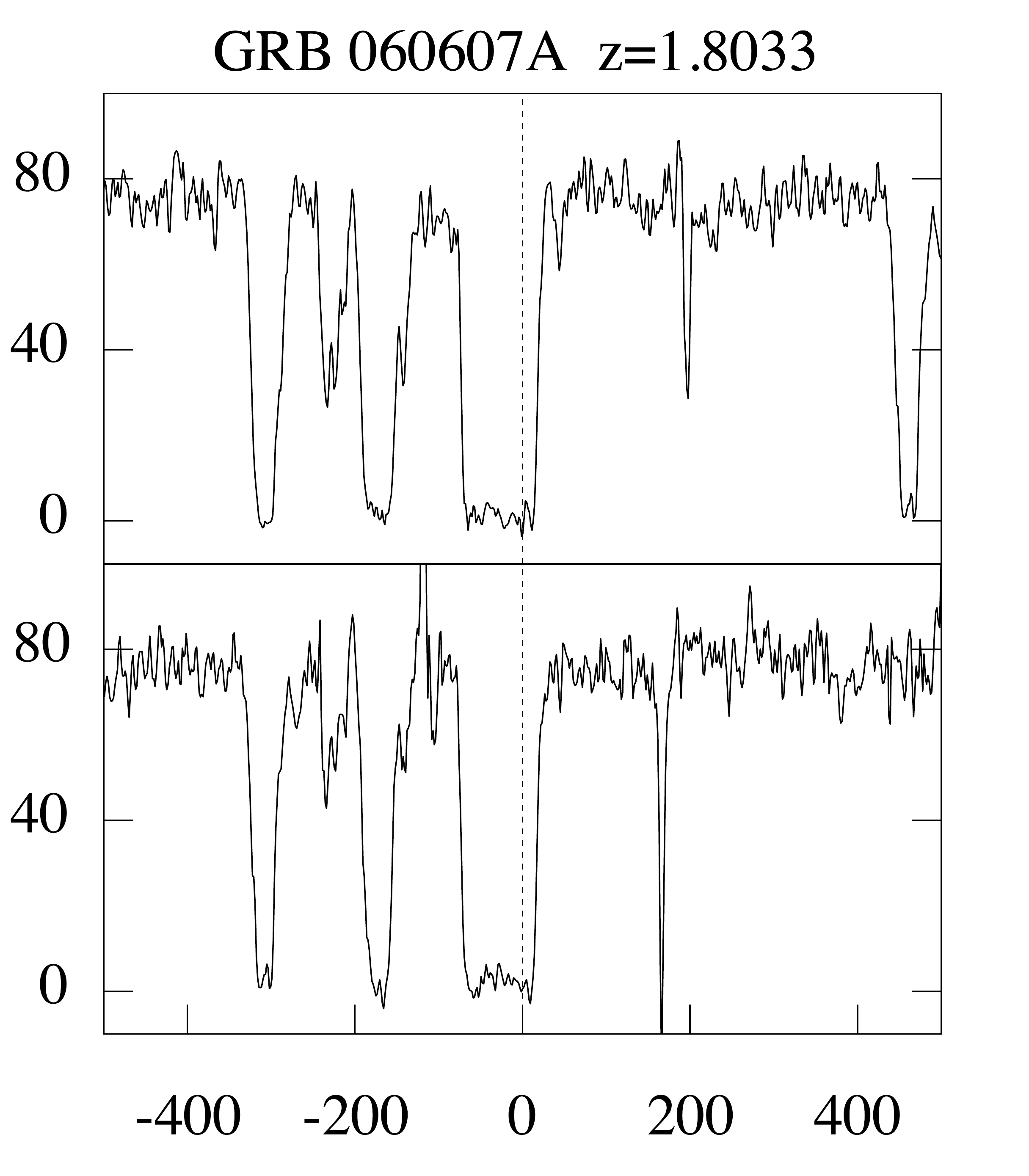}
\includegraphics[width=0.3\linewidth]{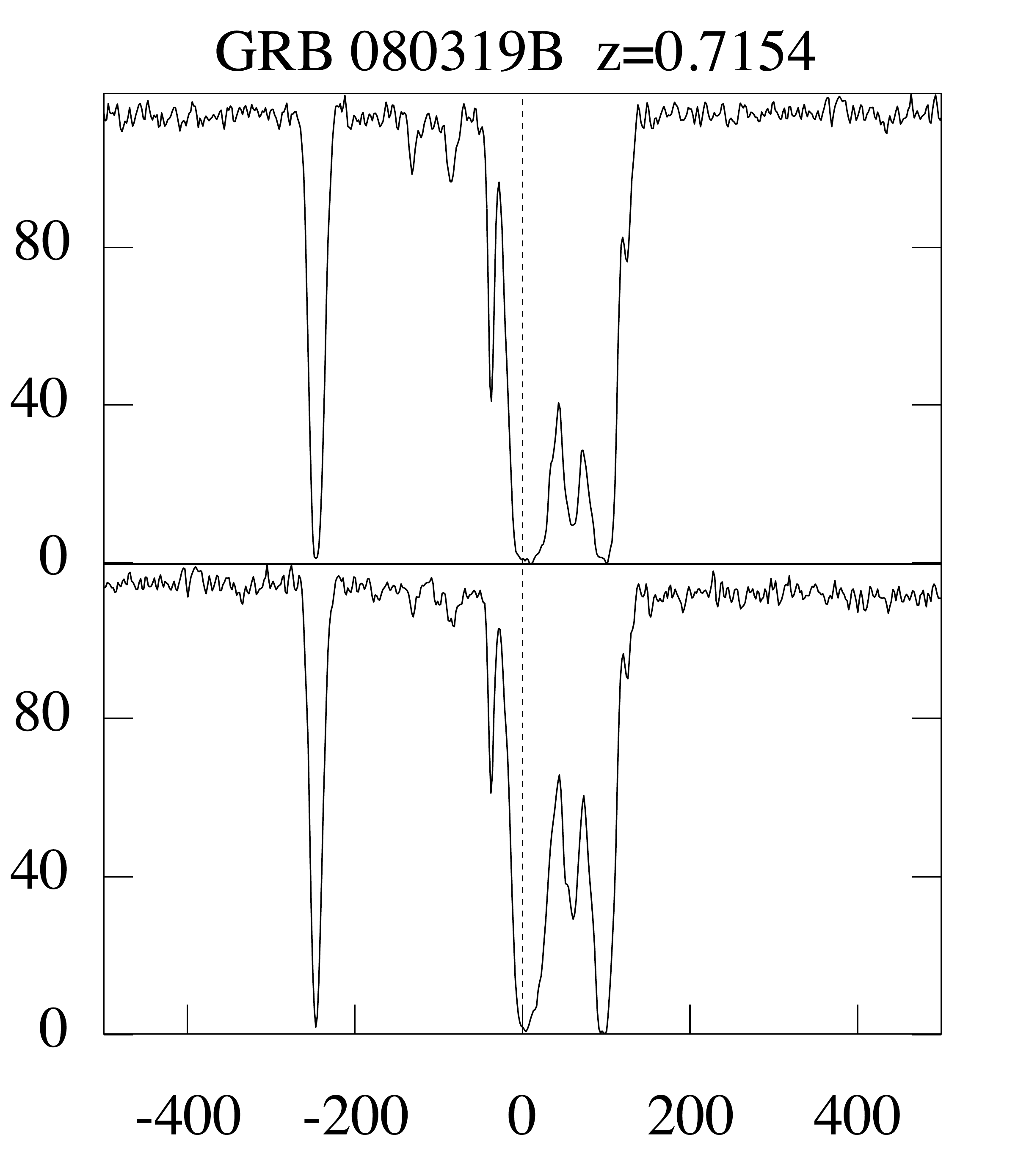}
\caption{The $W_{\rm r} >1.0$\,\AA \,Mg~{\sc ii} absorption systems detected in the UVES spectra. 
the Mg~{\sc ii} $\lambda$2796 (top) and $\lambda$2803 (bottom) absorption lines are displayed for each system on a relative velocity scale 
with 0 km s$^{-1}$ centered at the redshift reported in Table 1.}
\label{Mgstrong1}
\end{center}
\end{figure}

The profiles of the two transitions
of the  $W_{\rm r}$~$>$~1~\AA~ Mg~{\sc ii} systems found in the UVES 
spectra are plotted on a velocity scale in Fig.~\ref{Mgstrong1}. The profiles
are complex, spread over at least 200~km~s$^{-1}$ and up to $\sim600$~km~s$^{-1}$
and show a highly clumpy structure.

\begin{figure}[]
\begin{center}
\includegraphics[width=0.95\linewidth]{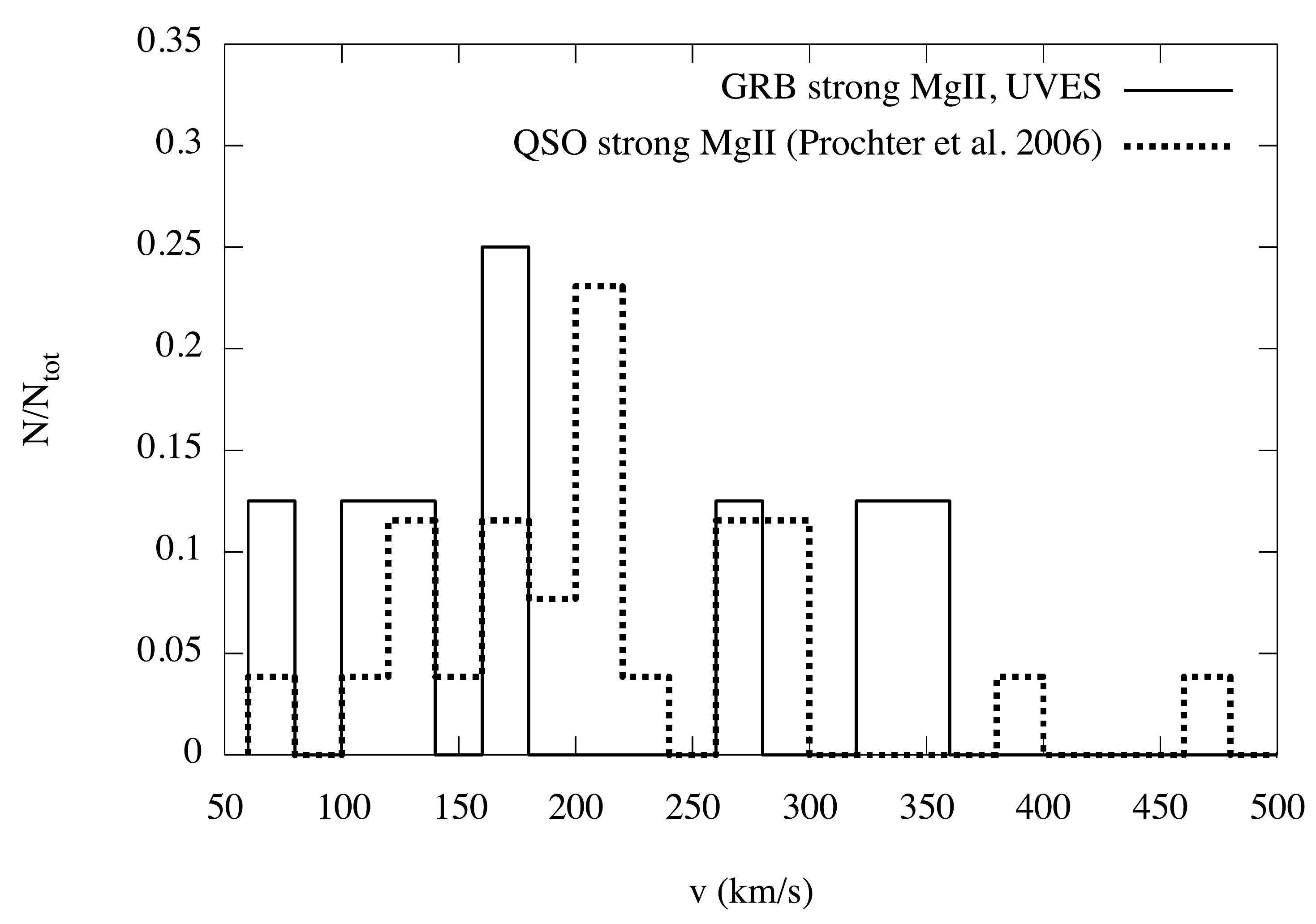}
\caption[ Distribution of the velocities of the strong Mg~{\sc ii} systems ]{The solid line histogram represents the distribution of the velocities of the strong Mg~{\sc ii} systems in the UVES 
sample calculated following \cite{Ledoux2006a}. The thick dashed histogram refers to the 27 SDSS Mg~{\sc ii}
systems observed at high spectral resolution \citep{Prochter2006}. 
}
\label{vdist}
\end{center}
\end{figure}

We measure the velocity spread of the Mg~{\sc ii} systems with $W_{\rm r}$~$>$~1.0~\AA~in the {\it UVES sample} 
following \cite{Ledoux2006a}. We therefore use a moderately saturated low ionization absorption line 
(e.g. FeII${\lambda2344}$, SiII${\lambda1526}$) and
measure the velocity difference between the points of the absorption profiles
at which 5\% and 95\% of the absorption occurs. 
This method is defined in order to measure the velocity width of the bulk of gas, avoiding contamination by satellite components which have negligible contribution to the total metal column density.
Note that using this definition implies that the measured velocity spread is usually smaller
than the spread of the Mg~{\sc ii} profile which is strongly saturated.
In case no moderately saturated absorption line is available, we use the mean value of
the velocity widths calculated both from a saturated line and an optically thin line.
Results are given in column 9 of Table~\ref{dla}.

The velocity-spread distribution is shown for UVES systems with $W_{\rm r}$~$>$~1~\AA~
in Fig.~\ref{vdist} together with the SDSS QSO distribution from 27 SDSS systems 
with $W_{\rm r}$~$>$~1~\AA~ observed at high spectral resolution \citep{Prochter2006}. 
Although the statistics are small, the UVES and SDSS distributions
are statistically similar. A KS test gives 65\% chance that the two samples are drawn from 
the same population (see also \citealt{Cucchiara2008a}).

\subsection{Dust extinction}

It has been proposed that a dust bias 
could possibly affect the statistics of strong Mg~{\sc ii} systems. Indeed, if part of the population
of strong Mg~{\sc ii} systems contains a substantial amount of dust, then the
corresponding lines of sight could be missed in QSO surveys because of the attenuation
of the quasar whereas GRBs being intrinsically brighter, the same lines of sight 
are not missed when observing GRBs. 
The existence of a dust extinction bias in DLA surveys is still a debated topic among the QSO community
although observations of radio selected QSO los \citep{Ellison2004} seem to show that, if any, this
effect is probably small (see also \citealt{Pontzen2008}).

\begin{table*}[]
\caption{Iron to Zinc or Silicon ratio and extinction estimate for 4 strong Mg~{\sc ii} systems.}
{ 
\begin{center}
\begin{tabular}{lccccccccc}
\hline\hline
\\[0.000001ex]
& $z$ & $N$(FeII) & $N$(ZnII) & $N$(SiII) & [Fe/Zn] & [Fe/Si] & $N_{\rm Fe}^{\rm dust}$ & $ A_{\rm V}$ \\  
&     &(cm$^{-2}$) & (cm$^{-2}$) &(cm$^{-2}$) &     &        & (cm$^{-2}$) & \\
\hline
\Huge
\\[0.05ex]
GRB021004&1.3800&$15.09\pm0.05$&&$15.19\pm0.09$&&$-0.05\pm0.10$&$14.18\pm0.23$&$\sim0$\\[1ex]
\hline
\\[0.05ex]
GRB021004&1.6026&$14.60\pm0.03$&$12.78\pm0.02$&&$-1.03\pm0.04$&&$15.59\pm0.02$&$<0.2$\\[1ex]
\hline
\\[0.05ex]
GRB060418&1.1070&$14.69\pm0.01$&$12.87\pm0.03$&&$-1.05\pm0.03$&&$15.68\pm0.03$&$<0.2$\\[1ex]
\hline
\\[0.05ex]
GRB060607A&1.8033&$14.07\pm0.03$&&$14.36\pm0.10$&&$-0.24\pm0.10$&$13.94\pm0.17$&$\sim0$\\[1ex]
\hline

\end{tabular}
\end{center}
}

\label{dust}
\end{table*}%

We can use our UVES lines of sight to estimate the dust content of strong Mg~{\sc ii} systems from
the depletion of iron compared to other non-depleted species as is usually done in DLAs.
We have therefore searched for both Fe~{\sc ii} and Zn~{\sc ii} absorption lines 
(or Si~{\sc ii}, in case the Zn~{\sc ii} lines are not available) associated to the strongest 
Mg~{\sc ii} systems present in the UVES spectra. Because the spectra do not always cover the
relevant wavelength range, the associated column densities could be measured only 
for 4 out of the 10 systems (see Table~\ref{dust}). We estimate the depletion factor, and therefore 
the presence of dust, from the metallicity ratio  
of iron (a species heavily depleted into dust grains in the ISM of our Galaxy) to zinc (that is little depleted), 
[Fe/Zn]~=~(Fe/Zn)$-$(Fe/Zn)$_{\sun}$ (or [Fe/Si]). We also determine the iron dust phase column 
density ($N_{\rm Fe}^{\rm dust}$) using the formula given by \cite{Vladilo2006} and from this
we infer an upper limit on the corresponding flux attenuation $A_{\rm V}$ from their Fig.\,4. 
We used also the estimate given by 
\citep{Bohlin1978,Prochaska2002}: 

\begin{equation}
A_{\rm V}= 0.5\frac{N({\rm HI})}{10^{21}}\kappa=0.5\frac{10^{\rm X}}{10^{21}\times 10^{\rm [X/H]_{\sun}}}(1-10^{\rm [Fe/X]})
\end{equation} 

with $\kappa=10^{\rm [X/H]}(1-10^{\rm [Fe/X]})$ representing the dust-to-gas ratio and X corresponding to Zn or Si if Zn is not available.

\begin{figure}[]
\begin{center}
\includegraphics[width=0.95\linewidth]{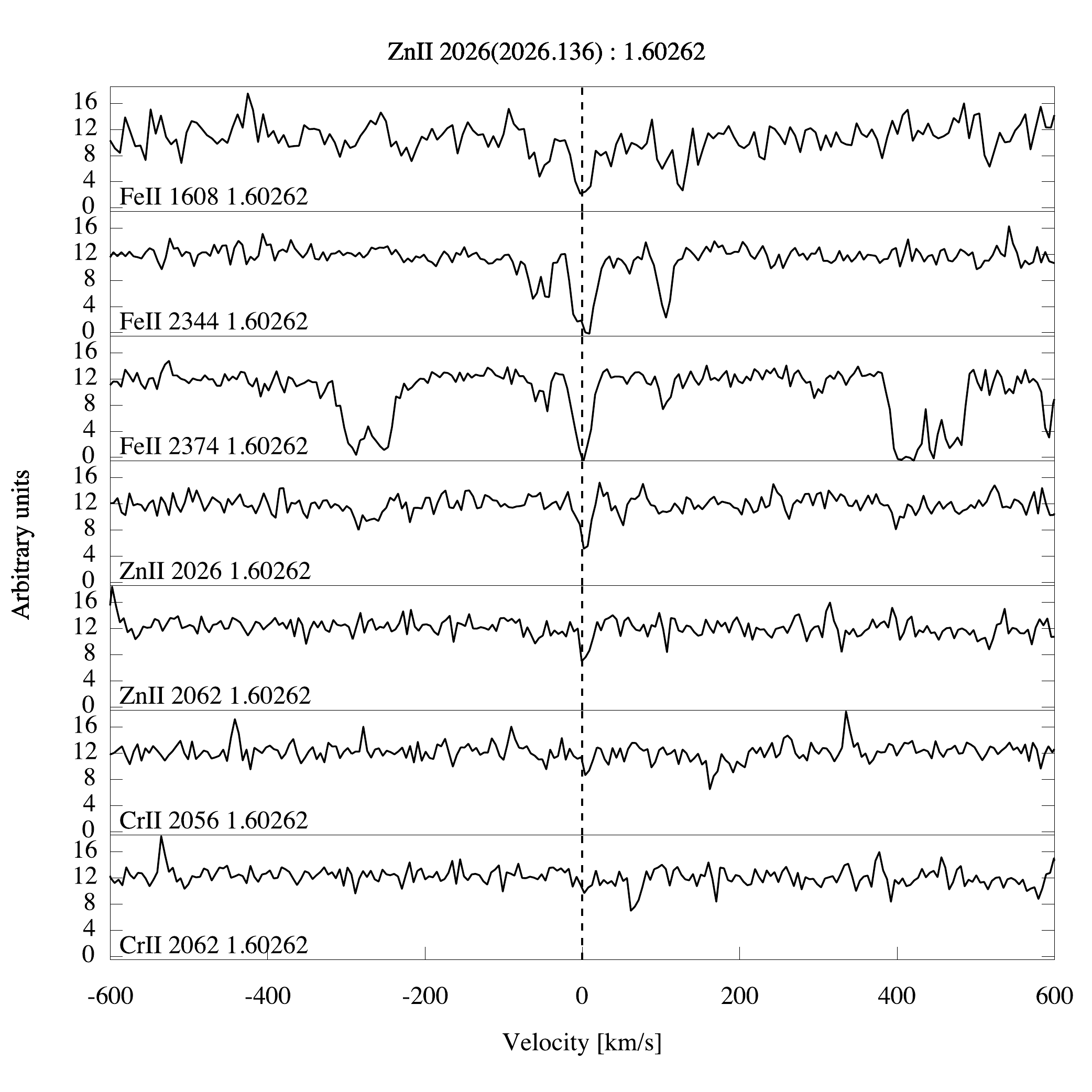}
\caption{ Fe\,II($\lambda$$\lambda$1608,2344,2374), Zn\,II($\lambda$$\lambda$2026,2062) and Cr\,II($\lambda$$\lambda$2056,2062) transition lines at $z_{\rm abs}$~=~1.6026 in the spectra of the afterglow of GRB\,021004, shown in velocity space. The dashed line correspond to $z_{\rm abs}$~=~1.6026. The Zn\,II absorption lines are clearly detected for the central component of the system.
}
\label{metal}
\end{center}
\end{figure}

Results are reported in Table~\ref{dust}. It can be seen that although depletion of iron 
can be significant, the corresponding attenuation is modest because the column densities
of metallic species are relatively small owing to low metallicities (see also Table~5).
Indeed we find high dust depletion in two systems at $z_{\rm abs}$~=~1.107 toward
GRB\,060418 (see also \citealt{Ellison2006}) and at $z_{\rm abs}$~=~1.6026 toward
GRB\,021004.
Their column densities are low however so that the inferred $A_{\rm V}$ does not exceed values 
typically found for QSO los (see \citealt{Vladilo2006} and \citealt{Prantzos2000}). \cite{Cucchiara2008a} do not detect the Zn\,II absorption lines for the systems at $z_{\rm abs}$~=~1.6026 toward
GRB\,021004. They report an upper limit of $W_{\rm r}<0.016$~\AA~ whereas we find $W_{\rm r}=0.07$~\AA. The detection of Zn\,II($\lambda$$\lambda$2026,2062) for the central 
component seems robust, as it is shown in Fig.\,\ref{metal}. In addition, the Cr\,II($\lambda$$\lambda$2056,2062) lines are also possibly detected.

We note that for GRB\,021004 and GRB\,060607A we could estimate the attenuation for 
ALL strong Mg~{\sc ii} systems identified along the los. Therefore our results do not show evidences to support the idea 
that a bias due to the presence of dust in
strong Mg~{\sc ii} systems could be an explanation for the overabundance of the strong Mg~{\sc ii} 
absorbers along GRB los. 
On the other hand, the fact that 2 out of 4 of the selected systems have a 
dust depletion higher than the values usually found for QSO DLAs \citep{Meiring2006} supports
the fact that these systems probably arise in the central part of massive halos where the
probability to find cold gas is expected to be higher.

\section{Estimating the H~{\sc i} column density}
\label{HIstrong}

\begin{table*}[]
\caption{Characteristics of Mg~{\sc ii} systems with $W_{\rm r}>1.0$\,\AA \, in the UVES sample.
}
{ 
\begin{center}
\small
\begin{tabular}{l c c c c c c c c c  }
\hline\hline\\
GRB & $z_{\rm abs}$  &  $W_{\rm r}^{MgII\lambda2796}$  & $W_{\rm r}^{MgI\lambda2852}$  &  $W_{\rm r}^{FeII\lambda2600}$  &  $N_{\rm ZnII}$  &  $N_{\rm SiII}$  &  $N_{\rm FeII}$  &  $^a \Delta v$  &  $^b \Delta v ^{MgII\lambda2796 }$\\[0.5ex]
    &  & (\AA) & (\AA) & (\AA) & cm$^{-2}$ & cm$^{-2}$ & cm$^{-2}$ & km s$^{-1}$ & km s$^{-1}$  \\
\hline
021004& 1.380  &  1.637  &  0.190  &  0.972  &    &  $15.19\pm0.09$  &$  15.09\pm0.05$  &  170  &89\\[0.5ex]
021004& 1.603  &  1.407  &  0.366  &  0.737  &  $12.78\pm0.02$  &    &  $14.60 \pm0.03$ &  164  &51\\[0.5ex]
050820A& 0.692  &  2.874  &      $N/A^c$     &        $N/A^c$        &    &    &           $N/A^c$                       &          & 228     \\[0.5ex]
050820A& 1.429  &  1.323  &  0.488  &  0.601  &    &    &  $14.34\pm0.03$  &  271  &98\\[0.5ex]
060418& 0.603  &  1.293  &  0.361  &  0.989  &    &   &  $16.43\pm0.04^d$  &  76  &130 \\[0.5ex]
060418& 0.656  &  1.033  &  0.078  &  0.486  &    &    &  $13.96\pm0.02$  &  136  &123 \\[0.5ex]
060418& 1.107  &  1.844  &  0.483  &  1.080  &  $12.87\pm0.03$  &    &  $14.69\pm0.01$  &  119  &221 \\[0.5ex]
060607A& 1.803  &  1.854  &  $>0.226$  &  0.825  &    &  $14.36\pm0.1$  &  $14.07\pm0.03$  &  333  &90\\[0.5ex]
080319B& 0.715  &  1.482  &  0.303  &  0.697  &    &    &  $14.00\pm0.01$  &  354  &142 \\[0.5ex]
 \hline
\end{tabular}
\end {center}
\footnotesize
$^a$ Velocities are measured following \cite{Ledoux2006a}; $^b$ Velocities are measured using the MgII$\lambda2796$ absorption considering only the central part of the system, following the method recommended by \cite{Ellison2009}; 
$^c$ Lines redshifted on top of the Ly${-\alpha}$ absorption associated to the GRB or in the Ly${-\alpha}$ forest; $^d$\,Saturated line.

 }
\label{dla}
\end{table*}

A key parameter to characterize an absorber 
is the corresponding H~{\sc i} column density. Unfortunately, the H~{\sc i}
Lyman-$\alpha$ absorption line of most of the systems is located below the atmospheric
cut-off and is unobservable from the ground.
If we want to characterize the systems with their H~{\sc i} column density or at least
an estimate of it, we have to infer it indirectly. 
For this we will assume that the strong systems seen in front of GRBs and QSOs are 
cosmological and drawn from the same population. We believe that the results discussed in the previous
section justify this assumption.

We use the velocity-metallicity correlation found by \cite{Ledoux2006a} to estimate the 
expected metallicity of the systems in the UVES sample with $W_{\rm r}$~$>$~1.0~\AA, assuming that the correlation is valid also for sub-DLA 
systems (P\'eroux et al. 2008). 
The relation was linearly extrapolated to the average redshift of the sample of systems with $W_{\rm r}$~$>$~1.0~\AA~, $\langle z \rangle = 1.11$, giving 
 [X/H]$=1.43{\rm log}\Delta v -3.78 $.
We infer the hydrogen column densities dividing the zinc, silicon or iron column densities 
measured in the UVES spectrum by the metallicity. 
Note that in most of the cases only Fe~{\sc ii} available (see Table~5). 
The $N$(H~{\sc i}) column density derived in these cases should be considered a lower limit because iron can
be depleted onto dust-grains. The results are shown in Table~\ref{dindex} (columns 3 and 4). 
The error on the procedure should be of the order of 0.5~dex (see \citealt{Ledoux2006a}).
We insist on the fact that
our aim is not to derive an exact H~{\sc i} column density for each system but rather to estimate
the overall nature of the systems.
We find that at least 3 of the 9 systems with $W_{\rm r}$~$>$~1~\AA~ could be DLAs. In any case, and even if we consider that we systematically overestimate
the column density by 0.5~dex, a large fraction of the systems should have log~$N$(H~{\sc i})~$>$~19.

Another method to infer the presence of DLAs among Mg~{\sc ii} absorbers 
is to use the `{\it D}-index' \citep{Ellison2009}.
We calculate the {\it D}-index for the systems in our sample following the recommended 
{\it D}-index definition by \cite{Ellison2009}
and using the formula:

 \begin{equation}
D=\frac{W_{\rm r}({\rm MgII}\lambda2796)({\rm \AA})}{\Delta v({\rm MgII}\lambda2796)({\rm km/s})}\times \frac{{\rm log}N({\rm FeII})}{15}\times1000 .
\label{eqd}
  \end{equation}

The width of the central part of the  Mg~{\sc ii}$\lambda2796$ absorption system,
$\Delta v$(MgII$\lambda$2796) in km/s, is reported in 
column 10 of Table~\ref{dla} while the resulting  {\it D}-index values are reported in 
Table~\ref{dindex} (column 5). Note that if we do not include the iron column density term in the {\it D}-index 
calculation, we obtain similar results.

We find that all the 9 systems have $D>7$. \cite{Ellison2009} find that $57^{+18.3}_{-14.0}\%$ of the systems 
having $D>7$ are DLAs. This results applied to our sample imply that at least 5 systems 
are DLAs.
We can compare the number of DLAs per unit redshift ($n_{\rm DLA}$) found for the QSO los by \cite{Rao2006} to that of our GRB sample.  
$n_{\rm DLA}$ is defined as the product of the Mg~{\sc ii} 
systems number density and the fraction of DLAs in a Mg~{\sc ii} sample. 
\cite{Rao2006} found 
$n_{\rm DLA, QSO}\sim0.1$ (with errors of the order of 0.02). 
If we assume that
three to five systems in our sample are DLAs
we find 
$n_{\rm DLA, GRB}\sim 0.22-0.36$ . This means that the number of DLAs is at least
two times larger along GRB lines of sight as compared to QSO lines of sight.

The above estimate of log~$N$(H~{\sc i}) can be considered as highly uncertain and the identification
of a few of the systems as DLAs can be questioned.

All this seems to indicate that GRBs favor lines of sight with 
an excess of DLAs. Since these systems
are more likely to be located in the central parts of massive halos, this may again favor the
idea that there exists a bias in GRB observations towards GRBs with afterglows
brighter because they are subject to some lensing amplification.

\begin{table}[h]
\caption{Inferred metallicity, $N_{\rm HI}$ and $D$-index of Mg~{\sc ii} systems with $W_{\rm r}>1.0$\,\AA \, in the UVES sample.
}
{ 
\small
\begin{tabular}{l c c c c   }
\hline
GRB & $z_{\rm abs}$   &   [X/H]$^a$ & $N_{\rm HI}({\rm cm}^{-2})^b$ &$D$-index$^c$\\

\hline
021004& 1.380   &$-0.63$& 20.3&$8.32\pm0.22$ \\[0.5ex]
021004& 1.603  &$-0.65$&20.8& $9.99\pm0.47$ \\[0.5ex]
050820A& 0.692  &      $$& &$9.0\pm0.03^d$  \\[0.5ex]
050820A& 1.429  &$-0.34$&  $>19.2$ &$8.13\pm0.22$ \\[0.5ex]
060418& 0.603  &$-1.13$&   $>22.0$&$10.39\pm0.09^e$ \\[0.5ex]
060418& 0.656    &$-0.77$&$>19.2$ &$7.51\pm0.05$\\[0.5ex]
060418& 1.107  &$-0.85$&21.1 &$8.17\pm0.06$ \\[0.5ex]
060607A& 1.803  &$-0.21$& 19.0&$9.12\pm0.06$ \\[0.5ex]
080319B& 0.715  &$-0.17$& $>18.7$&$8.06\pm0.01$ \\[0.5ex]
 \hline
\end{tabular}
\footnotesize

$^a$ Metallicities inferred using the velocity-metallicity 
relation found by \cite{Ledoux2006a}; $^b$ $N_{\rm HI}$ values inferred using the velocity-metallicity 
relation found by \cite{Ledoux2006a}. When only the iron column density is available, 
the $N_{\rm HI}$ value has to be considered has a lower limit due to possible dust extinction; 
$^c$ {\it D}-index calculated using Eq.\,\ref{eqd}, following the method recommended by \cite{Ellison2009};
$^d$  {\it D}-index calculated without including the iron column density term (see \citealt{Ellison2009});
$^e$ iron lines are saturated. The {\it D}-index calculated without including the iron column density term would be $D=9.48\pm0.08$;

 }
\label{dindex}
\end{table}

\section{Observed sub-DLA absorbers}
\label{DLAend}

\begin{figure}[htp]
\begin{center}
\includegraphics[width=0.95\linewidth]{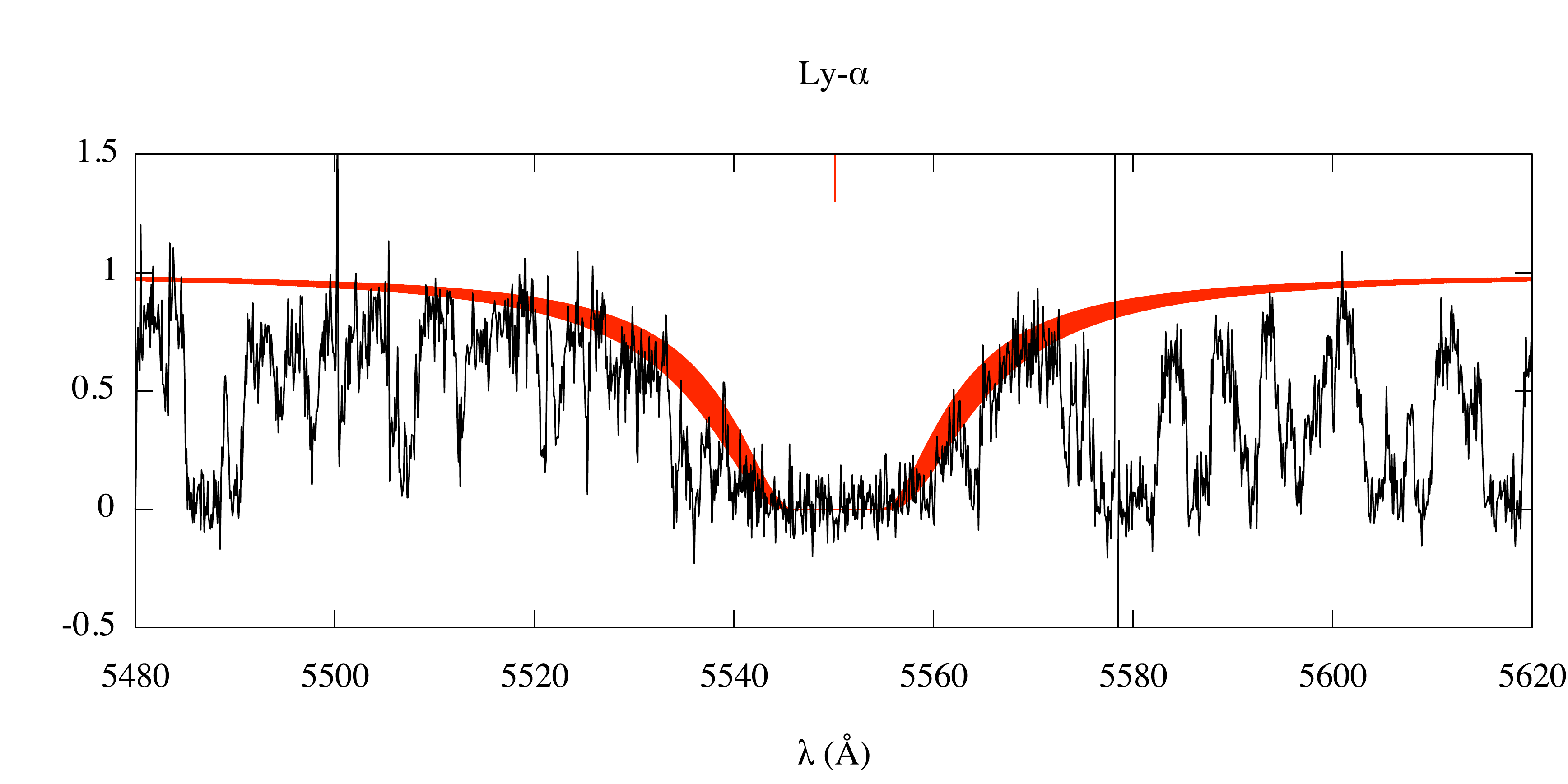}
\hfill
\includegraphics[width=0.95\linewidth]{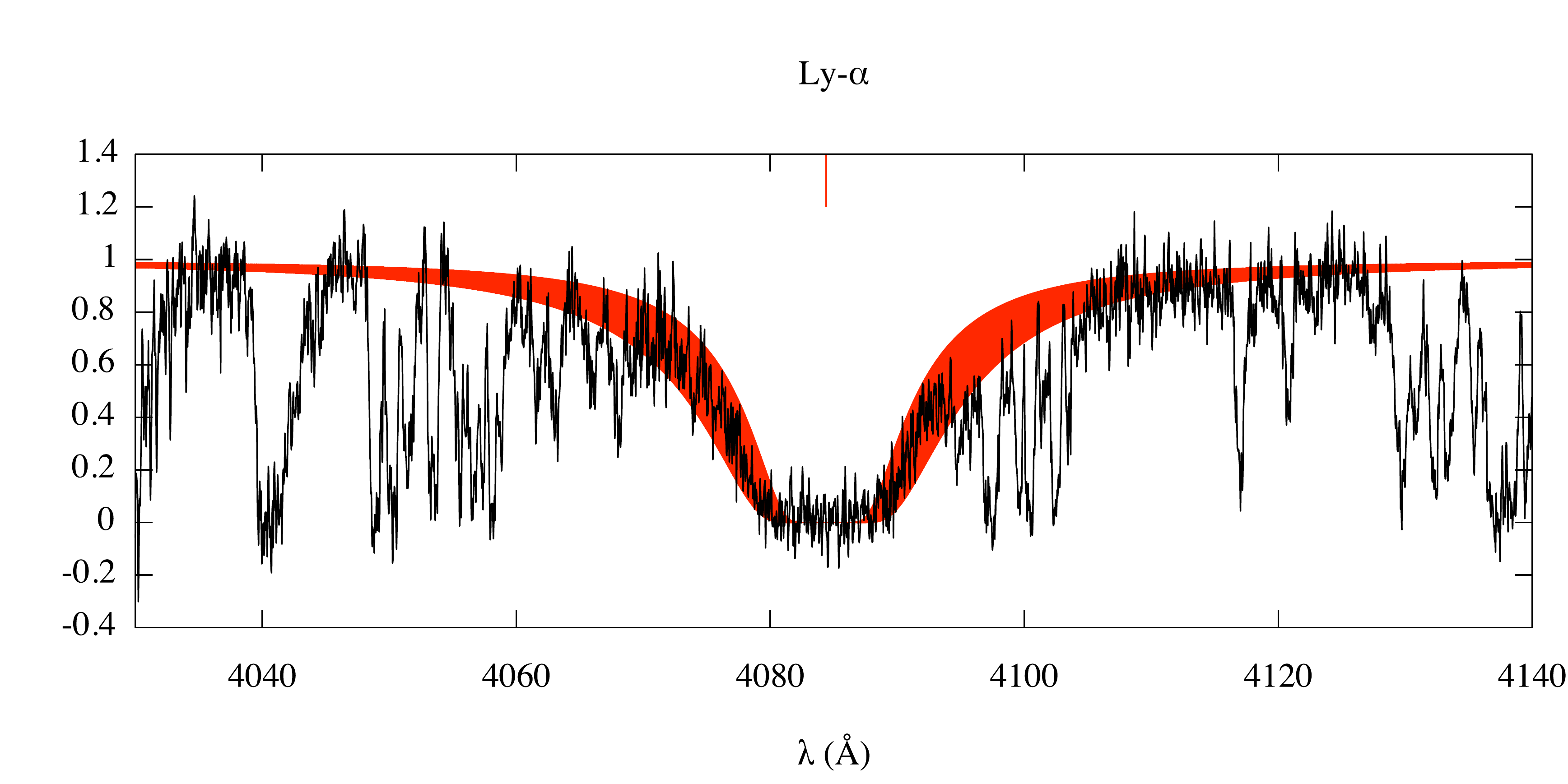}
\hfill
\includegraphics[width=0.95\linewidth]{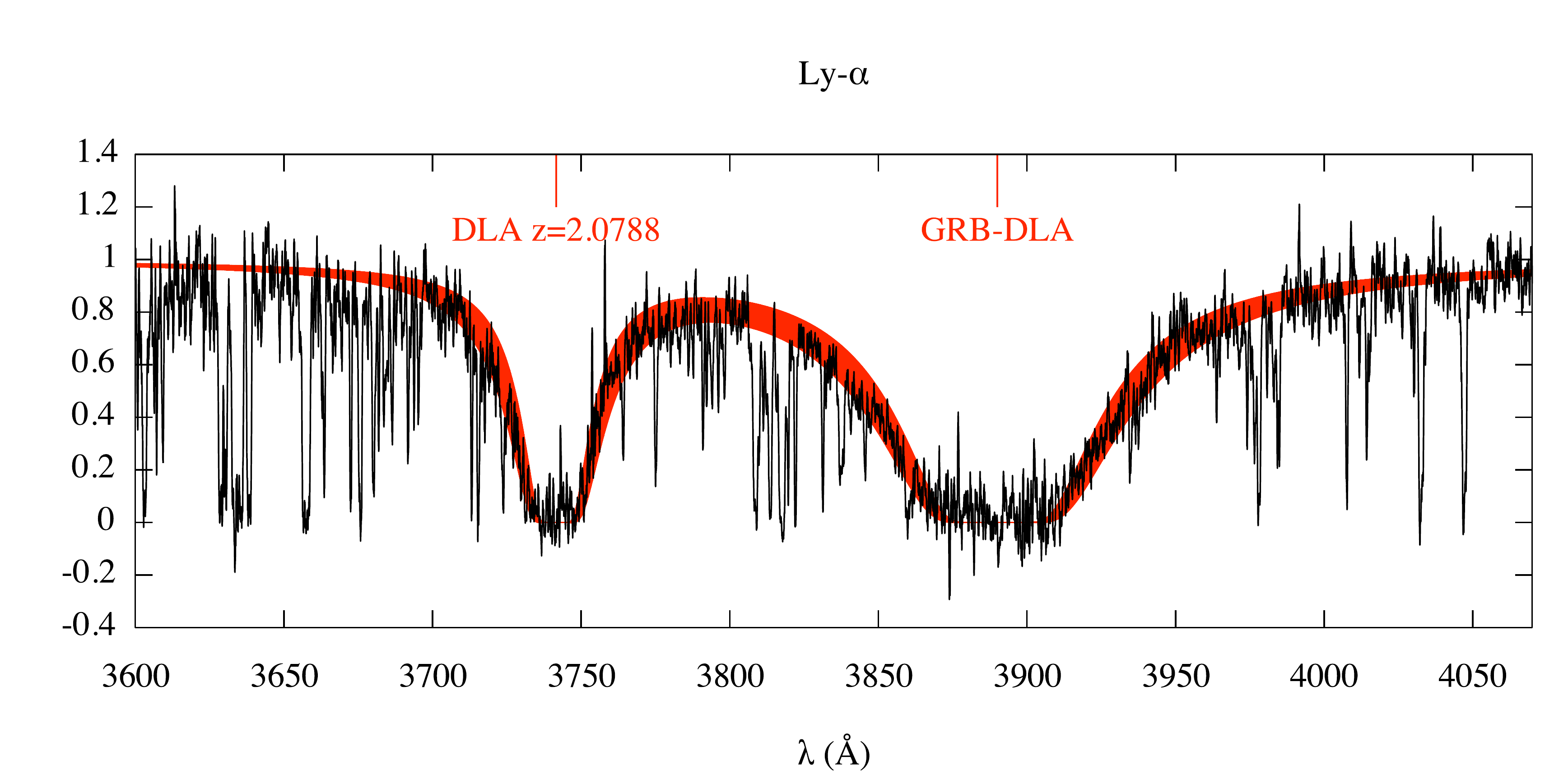}
\caption{Foreground (sub)-DLAs. {\it Top panel}: Sub-damped Lyman-$\alpha$ absorption 
at $z=3.5655$ toward GRB\,050730. {\it Middle panel}: Damped Lyman-$\alpha$ absorption 
at $z=2.3598$ toward GRB\,050820. {\it Bottom panel}: Sub-damped Lyman-$\alpha$ 
absorption at $z=2.0778$ toward GRB\,050922C (this system has been fitted together 
with the DLA associated to the GRB at $z=2.1999$). The red area correspond to the fit 
result covering the column density error range. }
\label{subfig}
\end{center}
\end{figure}

\begin{figure}[htp]
\begin{center}
\includegraphics[width=0.8\linewidth]{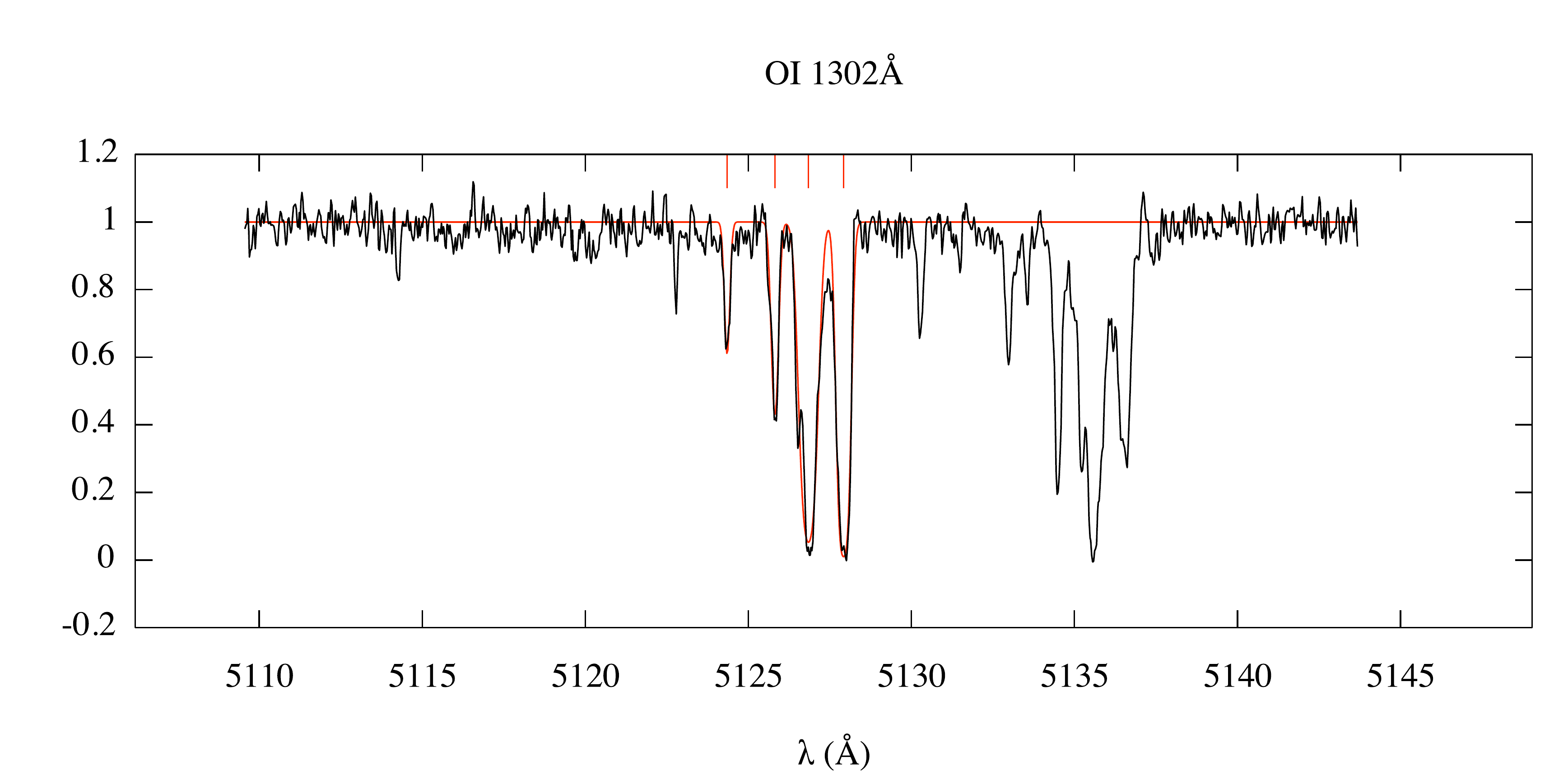}
\hfill
\includegraphics[width=0.8\linewidth]{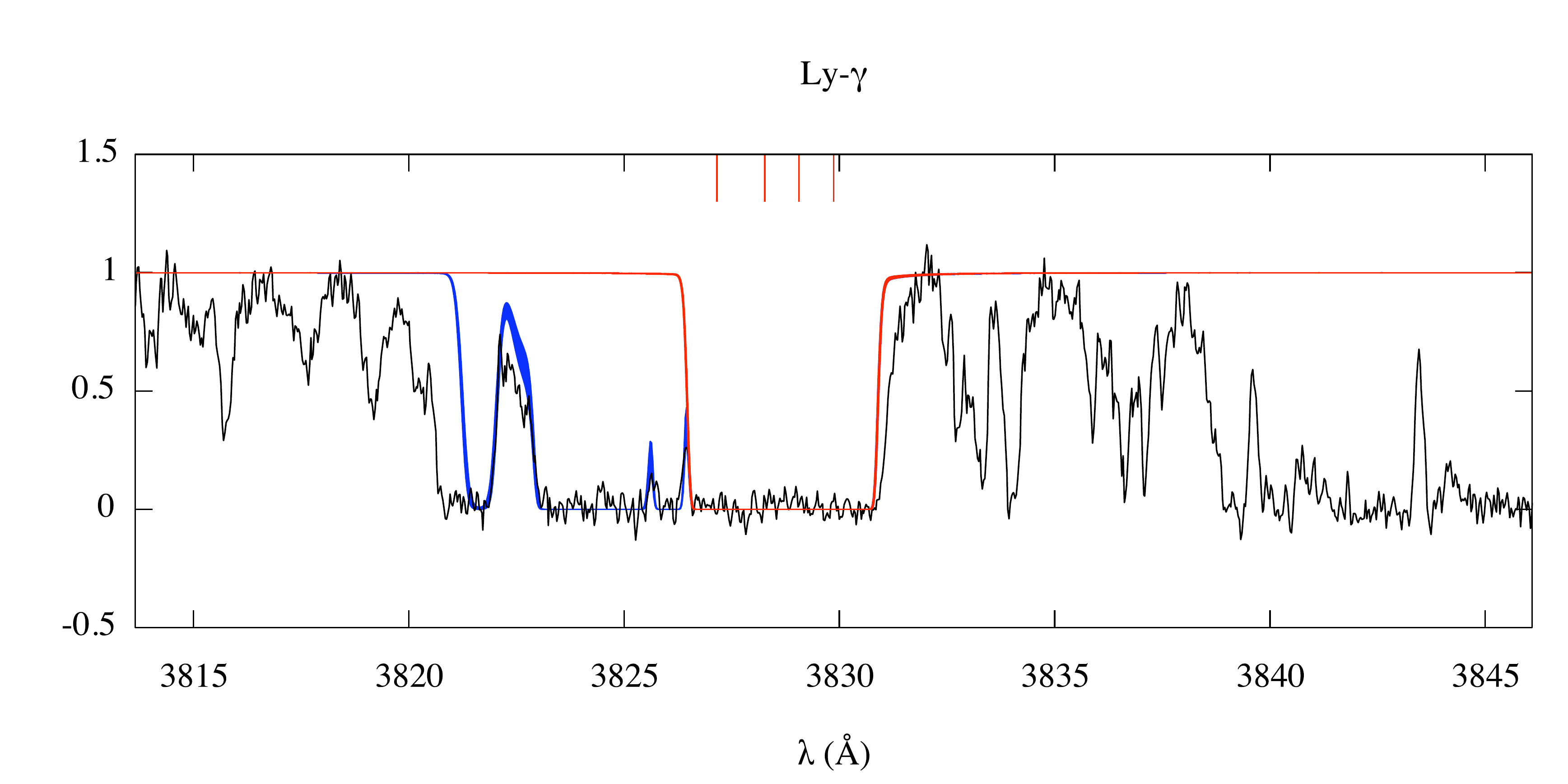}
\hfill
\includegraphics[width=0.8\linewidth]{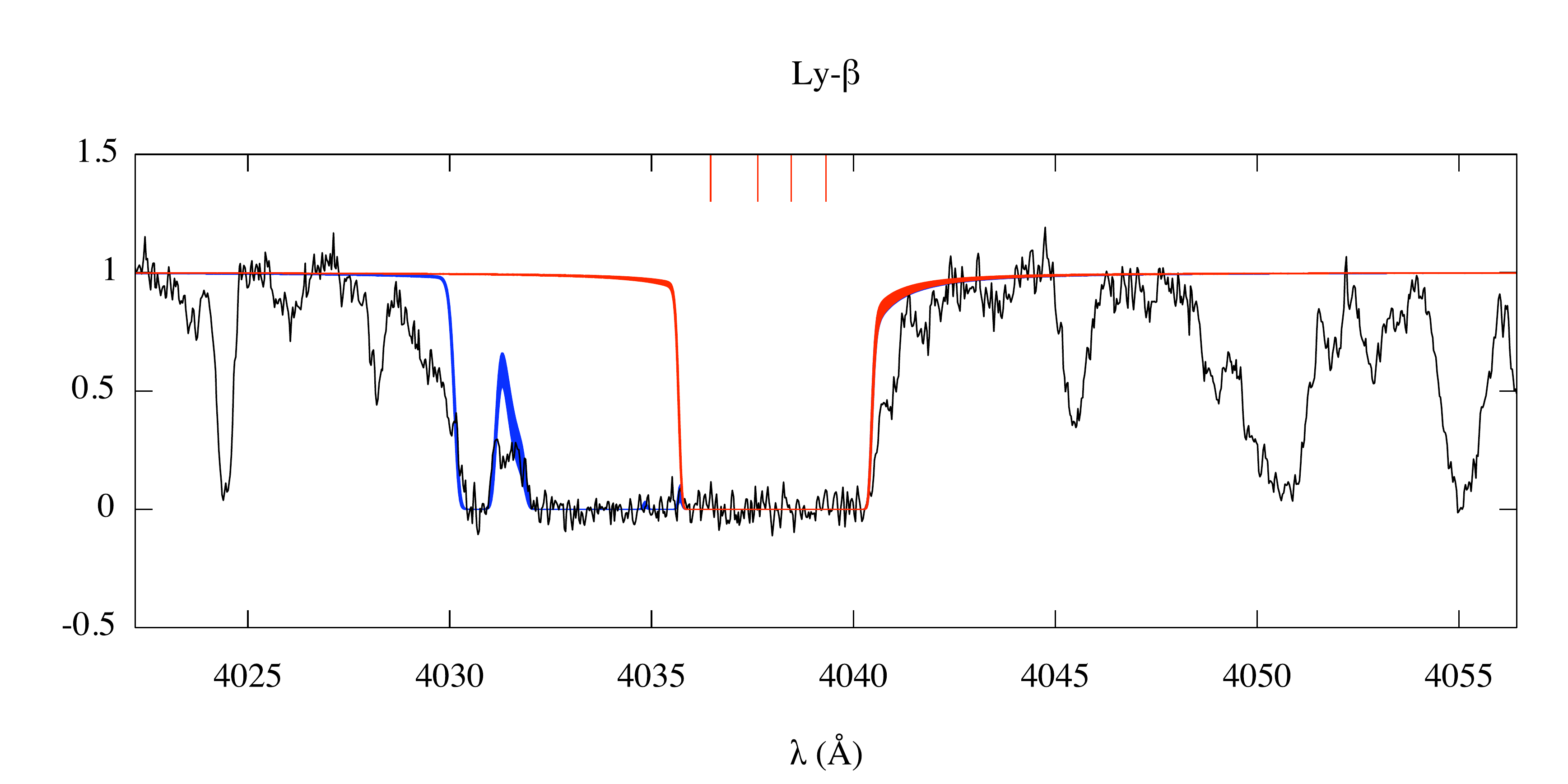}
\hfill
\includegraphics[width=0.8\linewidth]{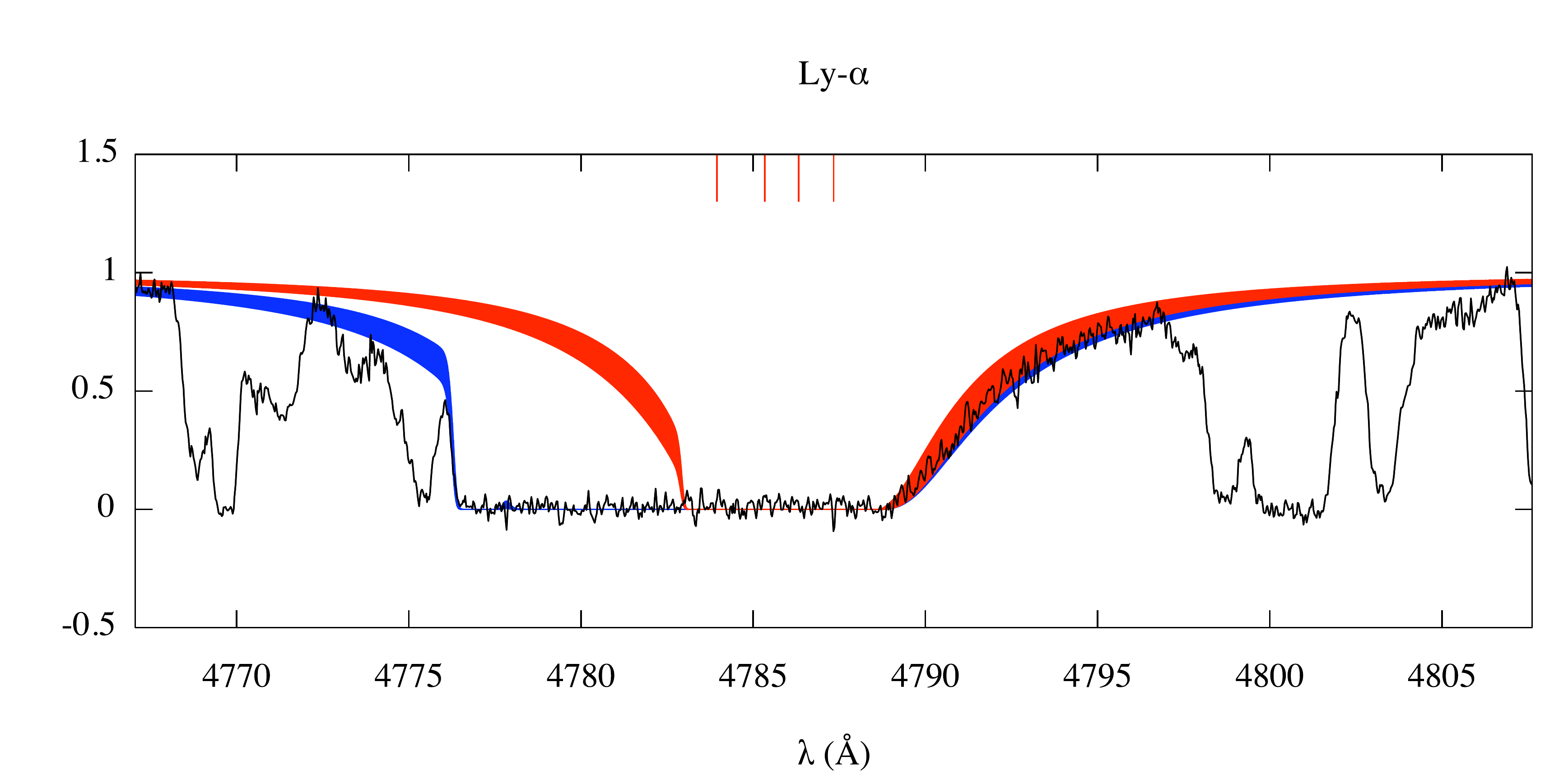}
\caption{The sub-DLA system at $z=2.9374$ along the los toward GRB\,060607A. The top panel 
shows the fit of the O~{\sc i}$\lambda$1302 absorption profile, while the following panels 
report the fit results of the Lyman-$\gamma$, Lyman-$\beta$ and Lyman-$\alpha$ absorption lines (from top 
to bottom, respectively). The red lines show the main $z=2.9374$ component (different lines include
errors) mostly constrained by the red wing of the Lyman-$\alpha$ absorption. 
In blue is shown the total fit including the sub-DLA component at $z=2.9322$. }
\label{ly060607}
\end{center}
\end{figure}

The UVES spectra often cover a substantial part of the Lyman-$\alpha$ forest
in front of the GRB. It is therefore possible to search directly for
strong H~{\sc i} Lyman-$\alpha$ absorption lines corresponding to
DLAs or, more generally, sub-DLAs. 
This can only be performed for 8 of the 10 GRBs in the sample: 
the redshifts of GRB\,060418 and GRB\,080319B are unfortunately
too low to allow the detection of the Lyman-$\alpha$ absorptions in the
UVES spectral range. We exclude from our search the DLA at the GRB redshift,
which is believed to be associated with the close surrounding of the GRB.
We find additional (sub-)DLAs (Fig.\,\ref{subfig}) along the los of GRB\,050730 (see also \citealt{Chen2005}), GRB\,050820, 
GRB\,050922C (see also \citealt{Piranomonte2008a}) and GRB\,060607A (Fig.\,\ref{ly060607}). 
The former three systems are simple, with a single Lyman-$\alpha$
component. The system detected at 2.9374 toward GRB\,060607A
is more complex and cannot be fit with a single component (sub-)\,DLA. 
The profile is made of two main clumps at $z=2.9322$ and 2.9374.
The latter is a blend of the four O~{\sc i} components 
(see top panel in Fig.~9). The constraints on the H~{\sc i} column densities
come mostly from the red wing of the Lyman-$\alpha$ line and
the structures seen in the Lyman-$\gamma$ line. 
We find that the two main H~{\sc i} components at $z=2.9374$ and 2.9322
have log $N$(H~{\sc i})~$=~19.4$ and $19.0\pm0.1$, respectively.
It is worth noting that the limit on the O~{\sc i} component
at $z=2.9322$ is log~$N$(O~{\sc i})~$< 12.6$ implying that the metallicity
in the cloud could be as low as [O/H]~$<$~$-2.9$ which would be
the lowest metallicity ever observed yet in such a system. Some ionization correction
could be necessary however in case the oxygen equilibrium is displaced
toward high excitation species by a hard ionizing spectrum.

Table\,\ref{Subtab} shows the column densities and abundances
of the systems assuming that all elements are in their neutral or singly ionized 
state. The values are similar to those usually found for
(sub)-DLAs along QSO los (see for example \citealt{Peroux2003}).

The presence of these systems allows us to extend the search for strong
systems along GRB lines of sight to higher redshift. 
The redshift paths (see Table\,\ref{Subtab}) have been
calculated considering $z_{\rm start}$ as the redshift of an absorber, for
which the corresponding Lyman-$\alpha$ line would be redshifted to the same wavelength
as the Lyman-$\beta$ line of the GRB afterglow. We therefore avoid confusion with
the Lyman-$\beta$ forest. 
$z_{\rm end}$ is fixed at 3000\,km/s from the GRB redshift.
The only exception to this rule is GRB\,050730 for which there is a
gap in the spectrum located at about 3000\,km/s from the GRB redshift and
starting at $\lambda=5096$~\AA. We use this wavelength to fix $z_{\rm max}$.
The corresponding total
redshift path is $\Delta z=4.17$. 
The sub-DLAs systems at $z=2.9322$ and 2.9374 along GRB\,060607A are separated by $\sim$400~km~s$^{-1}$, therefore 
for the statistical study we consider them
as a single
system. The resulting number density for systems with log $N$(H~{\sc i})$>$19.0 
is $0.96\pm0.48$ for an average redshift of
$\langle z \rangle =2.08$. At this redshift a value of
about $0.5\pm0.2$ is expected from QSO los \citep{Peroux2005}. There could be therefore 
an excess of such systems but the statistics is obviously poor.

It is intriguing however that the four systems we detect are all in the half 
of the Lyman-$\alpha$ forest closest to the GRB, with ejection velocities of about
25000, 22000, 12000 and 11000~km~s$^{-1}$ for GRB~050730, 050820A,
050922C and 060607A respectively. 
The corresponding ejection velocities found for the strong Mg~{\sc ii} systems 
(see column 7 of Table 2) are larger than 35000\,km/s, making an 
origin local to the GRB unlikely for the strong Mg~{\sc ii} systems. This fact therefore does not favor the 
possibility that the excess of strong Mg~{\sc ii} absorbers along GRB los is due to 
some ejected material present in the GRB environment. The much lower ejection velocities 
found for the sub-DLA absorbers may indicate that for these systems
the excess is not of the same origin as the excess of Mg~{\sc ii} systems
and that part of the gas in these systems could have been ejected by the GRB.

\section{Conclusion}
\label{Concl}

We have taken advantage of UVES observations of GRB afterglows obtained over the
past six years to build an homogeneous sample of lines of sight observed at high spectral resolution
and good signal-to-noise ratio (SNR$>$10). We used these observations
to re-investigate the claimed excess of Mg~{\sc ii} systems along GRB lines of sight
extending the study to smaller equivalent widths.
We also used these data to derive intrinsic physical properties of these systems.

Considering the redshift ranges $0.37<z<2.27$ of the SDSS survey used for the QSO statistics, we find an excess of strong intervening MgII systems along the 10 GRB lines of sight observed by UVES of a factor of $\sim 2$ compared to QSO lines of sigth. This excess is significant at $\sim 2\sigma$. Thanks to the quality of the UVES data it has been possible also to consider the statistics of the weak absorbers (0.3~$<$~$W_{\rm r}$~$<$~1.0~\AA). We find that the number of weak absorbers is similar along
GRBs and QSOs lines of sight.

\begin{table*}[htp]
\caption{Properties of the foreground damped Ly-$\alpha$ systems detected along the UVES GRB spectra. }

{ 
\scriptsize{
\begin{center}
\begin{tabular}{lc cccccc cccc }
\hline\hline
\\[0.000001ex]
&$\Delta z$$^a$& z & log $N_{\rm_{HI}}$&log $N_{\rm OI}$ &log $N_{\rm SiII}$  & log  $N_{\rm FeII}$  &  [Fe/Si]&[O/H]&[Si/H]&[Fe/H]&\\ 
&&(Sub-)DLA&($N$ in cm$^{-2}$)&($N$ in cm$^{-2}$)&($N$ in cm$^{-2}$)&($N$ in cm$^{-2}$)&&\\
\hline
\\[0.05ex]
GRB021004&0.487&\multicolumn{10}{l}{There are no intervening (Sub-)DLA along this los}\\[1ex]
\hline
\\[0.05ex]
GRB050730&0.542&3.5655&$20.2\pm0.1$&$N/A^b$&$<14.7$&$<13.6$&&&$<-1.0$&$<-2.3$\\[1ex]
\hline
\\[0.05ex]
GRB050820A&0.529&2.3598&$20.1\pm0.2$&$N/A^b$&$13.84\pm0.02$&$14.11\pm0.02$&$-0.22\pm0.02$&&$-1.5\pm0.2$&$-1.7\pm0.2$\\[1ex]
\hline
\\[0.05ex]
GRB050922C&0.468&2.0778&$20.65\pm0.15$&$>15.30$&$14.38\pm0.10$&$14.44\pm0.03$&$0.11\pm0.10$&$>-2.1$&$-1.83\pm0.18$&$-1.72\pm0.15$\\[1ex]
\hline
\\[0.05ex]
GRB060607A&0.596&$2.9374^c$&$19.4\pm0.1$&$>15.07$&$>14.74^d$&$14.16\pm0.03$&$<-0.53$&$>-1.1$&$>-0.2$&$-0.7\pm0.1$\\[1ex]
&&$2.9322^c$&$19.0\pm0.1$&$<12.8$&$<12.6$&$<12.6$&$ $&$<-2.9$&$<-1.9$&$<-1.9$\\[1ex]
\hline
\\[0.05ex]
GRB071031&0.540&\multicolumn{9}{l}{There are no intervening (Sub-)DLA along this los}\\[1ex]
\hline
\\[0.05ex]
GRB080310&0.501&\multicolumn{9}{l}{There are no intervening (Sub-)DLA along this los}\\[1ex]
\hline
\\[0.05ex]
GRB080413A&0.503&\multicolumn{9}{l}{There are no intervening (Sub-)DLA along this los}\\[1ex]
\hline

\end{tabular}
\end{center}
\smallskip

$^a$ Redshift path calculated from the Ly-$\beta$ GRB absorption line to 3000\,km/s from the GRB \lya, except for GRB\,050730 where the $z$ range ends at the beginning of the spectral gap at $\lambda=5096$\AA. 
$^b$ The line is blended with other lines in the Ly-$\alpha$ forest.
$^c$ To calculate the DLA number density we counted the $z=2.9374$ and $z=2.9322$ systems as a single system.\\
$^d$ Blended with the Si\,{\sc iv}$\lambda$1402 absorption line associated to the GRB.

}
}
\label{Subtab}
\end{table*}%

We increase the absorption length for strong systems to $\Delta z$~=~31.5
using intermediate and low resolution observations reported in the literature. The excess of strong MgII intervening systems of a factor of $\sim2$ is confirmed at $3\sigma$ significance. We therefore strenghten the evidence that the number of strong sytems is larger along GRB lines of sight, even if this excess is less than what has been claimed in previous studies based on a smaller path length  (see \citealt{Prochter2006a}). Our result and that of \cite{Prochter2006a} are different at less than $2\sigma$. The present result is statistically more significant. 

Possible explanations of this excess include: dust obscuration that could 
yield  such lines of sight to be missed in quasar studies; difference of
the beam sizes of the two types of background sources; gravitational lensing.
In order to retrieve more information to test these hypotheses we investigate in detail the properties of the strong Mg~{\sc ii} systems observed with UVES.
We find that the equivalent width distribution of Mg~{\sc ii} systems
is similar in GRBs and QSOs. This suggests that the absorbers are more extended
than the beam size of the sources which should not be the case for the different
beam sizes to play a role in explaining the excess \citep{Porciani2007}. 
In addition, we divide our sample in three redshift bins and we find that the number 
density of strong Mg~{\sc ii} systems is larger in the low redshift bins, favoring 
 the idea that current sample of GRB lines of sight
can be biased by gravitational lensing effect. We also estimate the dust extinction associated to the strong GRB 
Mg~{\sc ii} systems and we find that it is consistent with what is observed in standard (sub)-DLAs. It therefore
seems that the dust-bias explanation has little grounds.

We tentatively infer the H~{\sc i} column densities of the strong systems and we note that the number density of DLAs per unit redshift in the UVES 
sample is probably twice larger than what is expected from QSO sightlines.
As these sytems are expected to arise from the central part of massive haloes, 
this further supports the idea of a gravitational lensing amplification bias. This hypothesis could be also supported by the results recently obtained by \citep{Chen2008}. These authors analyzed 7 GRB fields and found the presence of at least one addition object at angular separation from the GRB afterglow position
in all the four fields of GRBs with known intervening strong
MgII galaxies. In contrast, none is seen at this small angular separation in
fields without known Mg II absorbers.

We searched the Lyman-$\alpha$ forest probed by the UVES spectra for the presence
of damped Lyman-$\alpha$ absorption lines. We found four sub-DLAs with log~$N$(H~{\sc i})~$>$~19.3
over a redshift range of $z=4.3$. This is again twice larger than what is expected
in QSOs. However the statistics is poor. It is intriguing that these
systems are all located in the half redshift range of the Lyman-$\alpha$ forest closest to
the GRB. It is therefore not excluded that part of this gas is somehow 
associated with the GRBs. In that case, ejection velocities of the order of 
10 to 25 000 km/s are required. 
 
\begin{acknowledgements}
S.D.V. thanks Robert Mochkovitch for suggesting that she applies to the Marie Curie EARA-EST program and the IAP for the warm hospitality.      
S.D.V. was supported during the early stage of this project by SFI grant 05/RFP/PHY0041 and the Marie Curie EARA-EST program. 
We are grateful to P. Noterdaeme and T. Vinci for precious help. 
\end{acknowledgements}

\bibliography{MgIIGRB}

\end{document}